\documentclass[aps,prb,twocolumn,showpacs,superscriptaddress]{revtex4-1}  
\usepackage{graphicx}  
\usepackage{dcolumn}   
\usepackage{bm}        
\usepackage{amssymb}   
\usepackage{amsmath}
\usepackage{braket}
\usepackage[colorlinks]{hyperref}

\begin{document}



\title{Environment tensor as order parameter for\\
 symmetry breaking and (symmetry-protected) topological orders
}
\author{Fangzhou Liu}
\affiliation{Department of Physics, Massachusetts Institute of Technology, Cambridge, Massachusetts 02139, USA}
\affiliation{Perimeter Institute for Theoretical Physics, Waterloo, Ontario, N2L 2Y5 Canada} 

\author{Xiao-Gang Wen}
\affiliation{Department of Physics, Massachusetts Institute of Technology, Cambridge, Massachusetts 02139, USA}
\affiliation{Perimeter Institute for Theoretical Physics, Waterloo, Ontario, N2L 2Y5 Canada} 
\affiliation{Institute for Advanced Study, Tsinghua University, Beijing 10084, China}

\date{\today}

\begin{abstract}
Spontaneous symmetry breaking is well understood through the classical
``Mexican Hat" picture, which describe many quantum phases of matter.
Recently, several new classes of  quantum phases of matter, such as topological
orders and symmetry protected topological (SPT) orders, were discovered.  In an
attempt to address the transitions between all those phases of quantum matter
under the same framework, we introduced an analogous yet very simple picture
for phase transitions in the context of tensor-networks.  Using a very simple
iteration process, we found that both symmetry breaking and some topological
phase transitions (for topological orders described by gauge theory and 1D SPT
orders) could be marked by a sudden change in the symmetry structure of the
so-called ``environment matrix''.  In this process, the environment matrix
serves as an ``order parameter'' that captures patterns of entanglement in
topological phases. The symmetry change in the environment matrix is very much
like the symmetry breaking of conventional order parameters.  We applied this
method to both the transverse Ising model ($1D$ and $2D$ honeycomb), spin-1
model ($1D$), and the Toric Code model in a magnetic field ($2D$ honeycomb),
and explored the corresponding symmetry structure changes in their environment
matrices in details.  With just a few variational parameters and a few minutes'
run time on a laptop, we could get the corresponding phase transition points
within a few percent error compared with the Quantum Monte Carlo results.
\end{abstract}

\maketitle


\section{Introduction}

In recent years, with the discoveries of quantum Hall
states\cite{KDP8094,TSG8259} and topological
insulators,\cite{KM0502,BZ0602,MB0706,FKM0703,K0986,RSF0957} the field of condensed
matter physics is focusing more and more on topological phases of matter. Lots
of progress has been made in the classification of topological
order\cite{W8987,WN9077,W9039,KW9327} in interacting bosonic/fermionic systems
through tensor network representation of many-body wave function and the
associated fixed-point tensors under wave function
renormalization,\cite{Levin_Strnt_05,Xie_LU_10,Liu_ModTrans_13,Gu_FermiLU_10}
which lead to tensor category theory of topological
order.\cite{Levin_Strnt_05,Xie_LU_10,KW1458,BBC1440} In the presence of
symmetry, tensor network and group cohomology\cite{Xie_SPT_12,K1459,W1477} also
lead to a classification of symmetry protected topological (SPT)
order.\cite{GU_CnrDoubleLine_09}

However, an important question is how to determine the topological order or SPT
order carried by a generic wave function or a generic tensor
network wave function.\cite{W9039,KW9327,W1221,ZGT1251,CV1223,ZMP1233,TZQ1251,Xie_LU_10,PT1241,HW1339,W1447,MW1418,HMW1457}  The tensors in
different generic tensor network can look similar, but represent different
topological/SPT prders.  This is because topological order is highly non-local.
All its features, including ground-state degeneracy, braidings and statistics
of the quasi-particles, topological entanglement entropy are global features.
One can only see those features after performing the wave function
renormalization.  This makes traditional theory of using local ``order
parameters'' to describe topological/SPT orders impossible.

Another difficulty to read topological/SPT orders from the local tensor is that
although the existing matrix-product representation has reached great success
in 1D, its higher-dimension extension is still a numerically formidable task,
and many $2D$ tensor-network renormalization scheme face the infamous ``corner
double-line'' problem,\cite{Gu_TnsrRG_08,GU_CnrDoubleLine_09} which
tensor-network renormalization quickly break down after a few iterations.
Thus a computationally efficient tensor-network method to implement RG is
badly desired.  It was in view of this that we developed our ``mean-field''
approach based on the environment matrix. 

This paper is structured as follows: 
In section \ref{1DMethod}, we first introduce the concept of
``environment matrix'' and outline how this method is applied 
in $1D$, with the example of the transverse Ising model.
In section \ref{SymmOfEnvr}, we make some detailed emphasis on the symmetry structure of the environment matrix, which leads to 
a characterization of different phases. 
In section \ref{MPSNoSymm} and \ref{1DSPT},
we detail how to detect different phases without knowing the
symmetry structure, which makes our method immediately applicable
to existing $1D$ numerical methods in identifying different SPT phases. 
Finally in section \ref{Sec2DIsing} and \ref{Sec2DTop}, we generalize this method to $2D$,
and apply it to both the transverse Ising model
and the Toric-Code model with a B-field on a honeycomb lattice.  

\section{1D environment tensor method}
\label{1DMethod}

\begin{figure}
\includegraphics[scale=0.8]{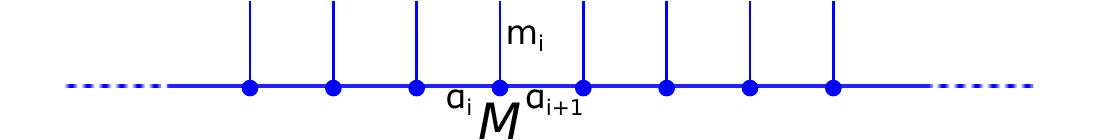}
\caption{\label{MPS} A matrix-product state. 
All the physical sites are represented by dots, and physical/internal
degrees of freedom by vertical/horizontal lines. }
\end{figure}

The environment tensor method has been widely applied in $1D$ 
systems through the study of Matrix-Product States (MPS).
\cite{Vidal_1D_07, Cirac_1D_09} 
Consider the $1D$ transverse Ising model on an infinite lattice:
\begin{align}
\label{1DIsing}
H= -J \sum_{<i,j>}\sigma_i^z \sigma_j^z - h \sum_{i}\sigma_i^x
\end{align}
where $\sigma$'s are the regular Pauli matrices.
Recall that the wave function of a $1D$ system could always be
written into a matrix-product form;
in particular, if the system is
translationally symmetric, then we have (in Figure \ref{MPS}):
\begin{align}
\label{1Dwavefunction}
\Ket{\Psi}=\sum_{ \{m_i\} }\sum_{ \{\alpha_i\} }
\text{Tr}[\prod_i M_{\alpha_i, \alpha_{i+1}}^{m_i} ] \Ket{ \{m_i\} }
\end{align}
where matrices $M$'s are independent of the site labels $i$ and are
labeled by the physical degrees of freedoms $m_i$.

\begin{figure}
\includegraphics[scale=0.7]{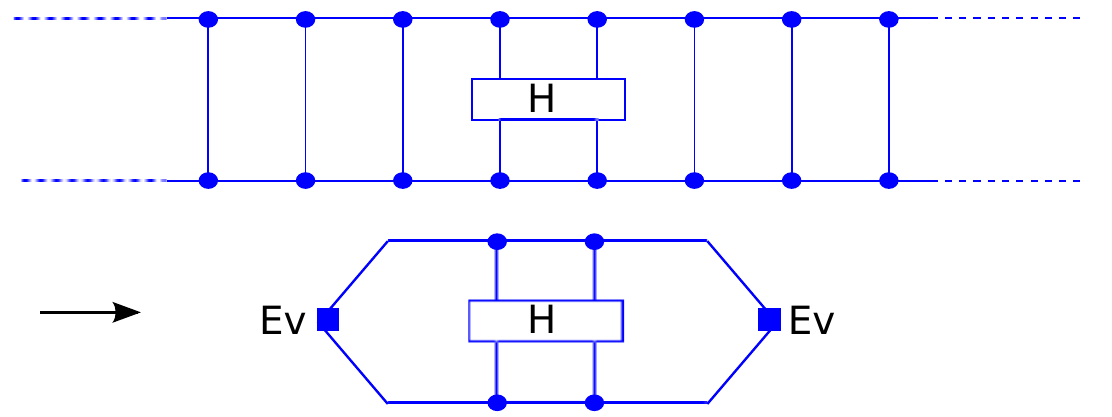}
\caption{\label{1D_Energy_infi} 
Approximate the energy through the environment matrix, $E$. 
Note that because the system is translation invariant,
we only need to calculate H for two neighboring sites.
In 1D, 
by assuming $M$ to be left-right symmetric, the environment matrices
on the left and right will be the same.}
\end{figure}

In $1D$, the environment tensor method is essentially a variational calculation
based on the above matrix-product state, with matrices $M$'s as variational
parameters.  We use the matrices $M$'s to obtain the average energy (see the
top of Fig. \ref{1D_Energy_infi}). We minimize the the average energy to obtain
$M$'s.

The calculation of average energy is actually a finite calculation.  The key is
to use ``environment matrix'' to capture the contributions from far-away sites.
This is graphically shown in the bottom of Fig. \ref{1D_Energy_infi}.  As can
be seen, there are two environment matrices, one on each side of the $1D$
chain.

So in the actual environment tensor method, we use the matrices $M$'s to obtain
the environment matrices $E$, and then use the matrices $M$'s and environment
matrices $E$'s to obtain the average energy.  We then minimize the average
energy to obtain $M$'s (and $E$'s).

\begin{figure}
\includegraphics[scale=0.6]{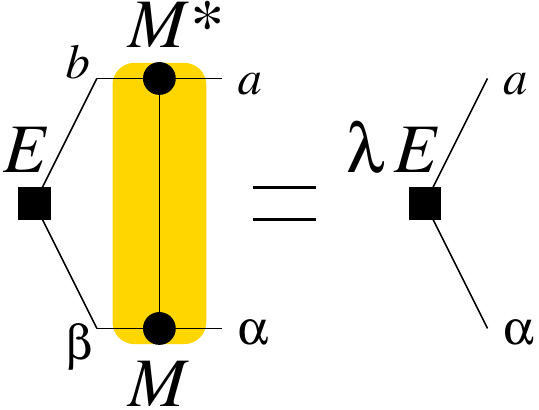}
\caption{\label{1Diter} A self-consistent condition 
that environment matrix  $E$ must satisfy, where
$\lambda$ is a scaling factor. 
The part enclosed by the shaded area is the so-called double-tensor
$T_{b\beta,a\alpha}$.}
\end{figure}

For fixed matrices $M$'s, the environment matrix could be obtained through
iterations.  As shown in Figure \ref{1Diter}, starting from some random initial
values $E_0$ that satisfies Tr$E_0^\dag E_0 =1$, we can update the environment
matrix using a ``double-tensor'', which is formed by two $M$ matrices with
physical indices contracted.  After applying the ``double-tensor'', $E_0$ is
changed to $\lambda_1 E_1$ where $E_1$ satisfies Tr$E_1^\dag E_1 =1$ and $\lambda_1$ is a
scaling factor.  After iterating enough number of times, a final stable
``environment matrix'' $E_\infty=E$ and a  final stable scaling factor
$\lambda_\infty = \lambda$ would be reached.  Note that this process, after viewing the
environment as a vector and the double-tensor as an operator, is essentially
equivalent to picking out the eigenvector with the largest absolute value of the eigenvalues of the
double tensor.  In this way, for each $M$, we can obtain the corresponding
environment matrix $E$ through iterations, and by applying $E$ both on the left
and on the right (see Figure \ref{1D_Energy_infi}), we can get the total
energy.  The variational calculation could then be carried out for different
values of $h/J$, and a phase diagram could then be obtained.

More specifically, we require our matrix-product state to have a $\mathbb{Z}_2$
symmetry that corresponds to spin up-down flipping, for both the
symmetry-breaking and the symmetric phases. So even in symmetry breaking phase,
we choose the ground state to be, say,
$(\ket{\uparrow\uparrow\dots}+\ket{\downarrow\downarrow\dots})$ when $h/J \to
0$.  Note that this is different from traditional symmetry breaking
description, where the ground state spontaneously picks one of the
ferromagnetic states.

Recall that on-site symmetry of the ground state requires matrices $M$'s to
transform in a special way w.r.t. symmetry:\cite{Cirac08}
\begin{align}
\label{MtrxSymm}
\sum_{m'}{g_{mm'}M^{m'}} = e^{i\theta_g} U^{\dagger}_g M^{m} U_g
\end{align}
Here, $m$ is the spin index, matrix $g_{mm'}$ represents the on-site symmetry
and acts on the spin basis, $\theta$ is a phase factor (set to $0$ in this
paper), $U_g$ is a unitary matrix acting on internal degrees of freedom, and
forms a projective representation of the symmetry group
$g$.\cite{PBT1039,CGW1107,SPC1139}

For internal dimension $D$ (dimension of $M$) being 2, we can choose $U_g =
\begin{pmatrix}1 & 0 \\ 0 & -1 \end{pmatrix}$,
then equation \eqref{MtrxSymm} reduces to
\begin{align}
\label{MtrxSymm1D}
M^{\uparrow} = U^{\dagger}_g M^{\downarrow} U_g,
\end{align}
and we thus have: $ M^{\uparrow} = \begin{pmatrix} a & b \\ b & c
\end{pmatrix}$ and $M^{\downarrow} = \begin{pmatrix} a & -b \\ -b & c
\end{pmatrix}$.  Here $M$'s are symmetric because of left-right symmetry, and
$a,b,c$ are free variational parameters.

\begin{figure}
\includegraphics[scale=1.0]{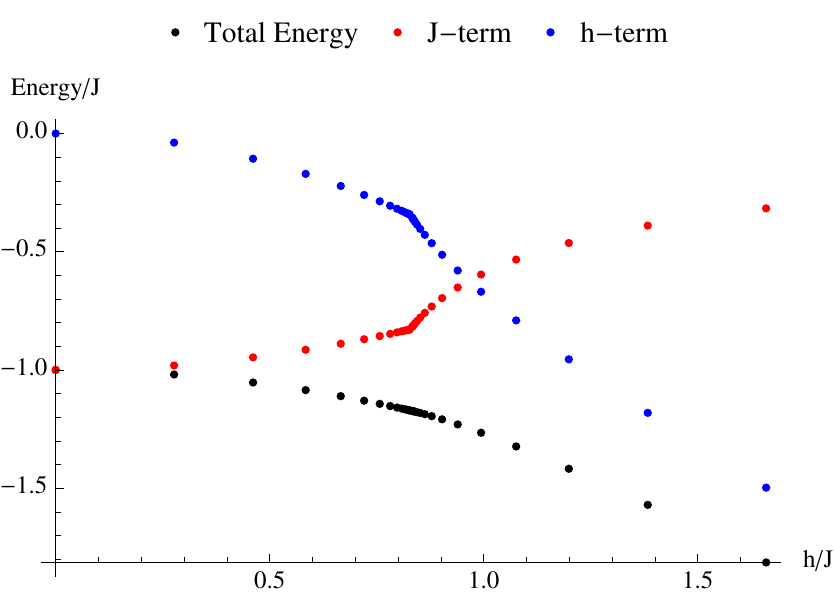}
\caption{\label{1DIsingEnergy} Energy as a function of h/J. The h- and J- terms
are also individually plotted in the graph, so the phase transition could be
easily spotted at h/J = 0.83. This result is obtained when internal dimension D=2,
and we've chosen the grid so that sample points are denser 
close to the transition point.}
\end{figure}

\begin{figure}
\includegraphics[scale=1.0]{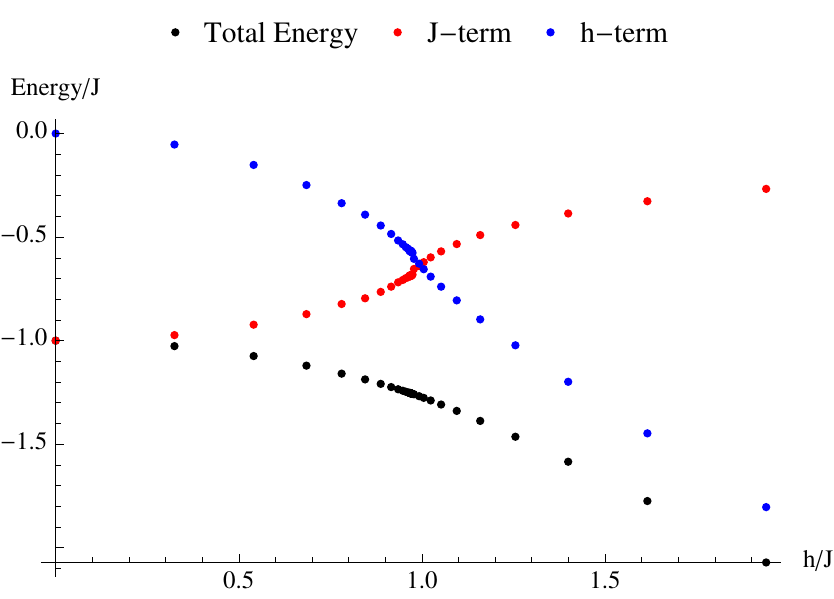}
\caption{\label{1DIsingEnergyD=4} Energy plot when internal dimension
increased to D=4. We can see the phase transition point occured around
h/J = 0.97, 
a big improvement from Fig. \ref{1DIsingEnergy}.}
\end{figure}

With the above symmetry analysis in mind, numerical simulation
could be run on our
$1D$ Ising model.  Following the previous discussion, for each $h/J \in
[0,\infty]$ in equation \eqref{1DIsing}, we minimize the energy by varying
$M$'s satisfying equation \eqref{MtrxSymm1D}.  By
plotting the two energy terms, a phase diagram is obtained
(see Fig. \ref{1DIsingEnergy}). 
For internal dimension
$D=2$, the phase transition occurred at $h/J = 0.83$, with an error of
$17\%$. 

We could easily improve the result by increasing the internal dimension.
For internal dimension $D=4$, we can choose $U_g$
in \eqref{MtrxSymm1D} to be 
$\begin{pmatrix}I & 0 \\ 0 & -I \end{pmatrix}$, where $I$ is the $2\times2$
Identity matrix. The most general symmetric $M$ satisfying 
\eqref{MtrxSymm1D} has $10$ variational parameters. Following
the same variational procedure, we can get the energy plot shown
in Fig. \ref{1DIsingEnergyD=4}.
The phase transition occurred at
$h/J = 0.97$, with a mere $3\%$ error. 

Note that in both calculations, we used symmetric matrices $M$'s
with all real parameters.  The typical runtime on a laptop was just a few
seconds in both cases.

\section{Symmetry structure of the environment matrix}
\label{SymmOfEnvr}

From the above plot, we see that there is a phase transition at $h/J\approx
0.83$.  To understand the phases on the two sides of the transition, let us
choose another basis
\begin{align}
\label{MPSAltBasis}
 \tilde M^{\uparrow} = W M^{\uparrow} W^\dag =
\frac 12 \begin{pmatrix} a+c+2b & a-c \\ a-c & a+c-2b \end{pmatrix}
\nonumber\\
 \tilde M^{\downarrow} = W M^{\downarrow} W^\dag =
\frac 12 \begin{pmatrix} a+c-2b & a-c \\ a-c & a+c+2b \end{pmatrix}
\end{align}
where $W= 2^{-1/2} \begin{pmatrix} 1 & 1 \\ 1 & -1
\end{pmatrix}$.
In the new basis the meaning of the $\tilde M$'s is more clear. 

When $b=0$, $\tilde M^{\uparrow}= \tilde M^{\downarrow}$, and the MPS is
a pure product state
$\otimes (|\uparrow\rangle +|\downarrow\rangle )$
that does not break the $Z_2$ symmetry. When $b\neq 0$ and $a=c$,
the MPS is a symmetry breaking state of the form $|\Psi\rangle + U
|\Psi\rangle$ where $U$ is $Z_2$ symmetry transformation.  But when  $b\neq 0$
and $a\neq c$, what is the nature of the MPS?

To answer such a question, we would like to study the symmetry structure of the
environment matrix $E$. 
We find that, depends on which phase we are in, the symmetry structure of $E$
will be very different.

As we mentioned before, the environment matrix $E$ and the associated scaling
factor $\lambda$ is calculated via the iteration (or the self consistent
condition) in Fig. \ref{1Diter}.  In general,
there can be many  environment matrices $E$ that satisfy the self consistent
condition. Here we choose those with largest absolute value of the scaling factor $\lambda$.  If
there are many  environment matrices with the degenerate largest absolute value of the scaling
factor, we then choose the  environment matrices with minimal ``entropy''
\begin{align} 
\label{MinEntropy}
S=\sum_i -s_i\ln s_i, 
\end{align} 
where $s_i$ is the singular
values of $E$.  This will give us a set of environment matrices $\{E\}$.

Next, we want to point out that environment matrix has only internal indices,
so for $E$, the symmetry transformation \eqref{MtrxSymm} translates to:
\begin{align}
\label{EnvrSymm}
E \to U^{\dagger}_g \cdot E \cdot U_g.
\end{align} 
If the $M$'s are invariant under the symmetry transformation (\ref{MtrxSymm}),
then the set of  environment matrices $\{E\}$ will be invariant under the above
transformation (\ref{EnvrSymm}).

If the action of  transformation (\ref{EnvrSymm}) is trivial on the set of
environment matrices $\{E\}$, then the MPS does not have symmetry breaking.  If
the action is non-trivial ({\it i.e.} generate a permutation of the set
$\{E\}$), then the MPS, in general, has a symmetry breaking; but this is not
guaranteed.

\begin{figure}
\includegraphics[scale=0.6]{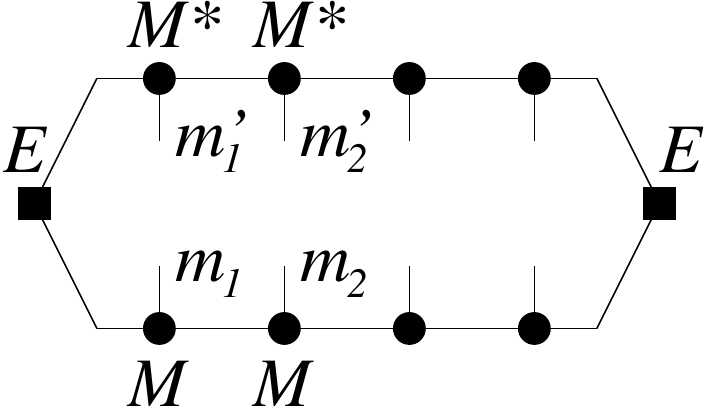}
\caption{\label{rho} 
The entanglement density matrix can be calculated from the environment matrix.
(The correct  entanglement density matrix should be calculated from the total
the environment tensor $E^0\otimes E^0 +E^1\otimes E^1$.)
}
\end{figure}

The reason for the complication is that there are zero-measure possibilities
that some internal bond degrees of freedom completely decouple from the
physical degrees of freedom.  To fix this problem, we may consider the
entanglement density matrix $\rho_{m_1m_2\cdots,m_1'm_2'\cdots}$ defined in
Fig. \ref{rho}.  We say two environment matrices are equivalent if they generate
the same $\rho_{m_1m_2\cdots,m_1'm_2'\cdots}$.  Let us use $\{E\}/\sim$ to
denote the equivalent class of the environment matrices. Then if the action
(\ref{EnvrSymm}) is non-trivial on $\{E\}/\sim$, then the MPS has a symmetry
breaking.

\begin{figure}
\includegraphics[scale=0.52]{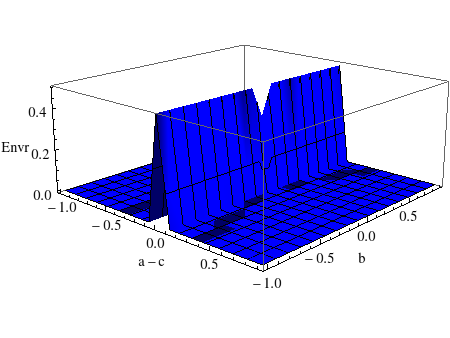}
\caption{\label{EnvrVSM} 
The off-diagonal term in environment matrix is plotted here
as a function of $a-c$ and $b$. We can see
the delta-function-like behavior when $a=c$. Note the dip
at $a=c$ and $b=0$: this point corresponds to the four-fold
degeneracy in the double-tensor, which is in the symmetric phase.
}
\end{figure}

With the above general discussion, we now go to our numerical results for
internal dimension $D=2$. 
For the symmetric phase, using the iteration method in Figure \ref{1Diter} and
after energy minimization, we obtain a final $E = \begin{pmatrix} s & 0 \\ 0 &
t \end{pmatrix}$, which gives an invariant $E$ under eqn
\eqref{EnvrSymm}.
As a result, the total environment tensor has a pure tensor product form
\begin{align}
\label{ESymm}
E^{tot} = E \otimes E,
\end{align}

For the symmetry-breaking phase, after minimizing entropy according
to eqn. \eqref{MinEntropy},
depending on the initial values of $E$, the iteration method
would give either  $E^{g_1} = \begin{pmatrix} p & p \\ p & p \end{pmatrix}$
or  $E^{g_2} = \begin{pmatrix} p & -p \\ -p &  p \end{pmatrix}$,
which transforms into each other under eqn \eqref{EnvrSymm}.
Both of these correspond to environment matrix of the fixed-point
wavefunction, as explained later in this section.
If we construct the total environment tensor
\begin{align}
\label{EnvrBrk}
E^{tot} = E^{g_1} \otimes E^{g_1} + E^{g_2} \otimes E^{g_2},
\end{align}
then the $\mathbb{Z}_2$ symmetry is restored, but now the
the total environment tensor does not have a pure tensor product form.

We are now in the position to answer the question raised at the begining
of this section. We now know that symmetry-breaking phase is signatured
by a non-zero off-diagonal term in the environment matrix. 
As shown in Fig. \ref{EnvrVSM}, if we plot
this off-diagonal term as a function of $a-c$ and $b$ in $M$ (see eqn \eqref{MPSAltBasis}),
then we see that the system is only in symmetry breaking
state when $a=c$ and $b\ne 0$. So when $b\neq 0$ and $a\neq c$, the state is in the
symmetric phase.

\begin{figure}
\includegraphics[scale=1.0]{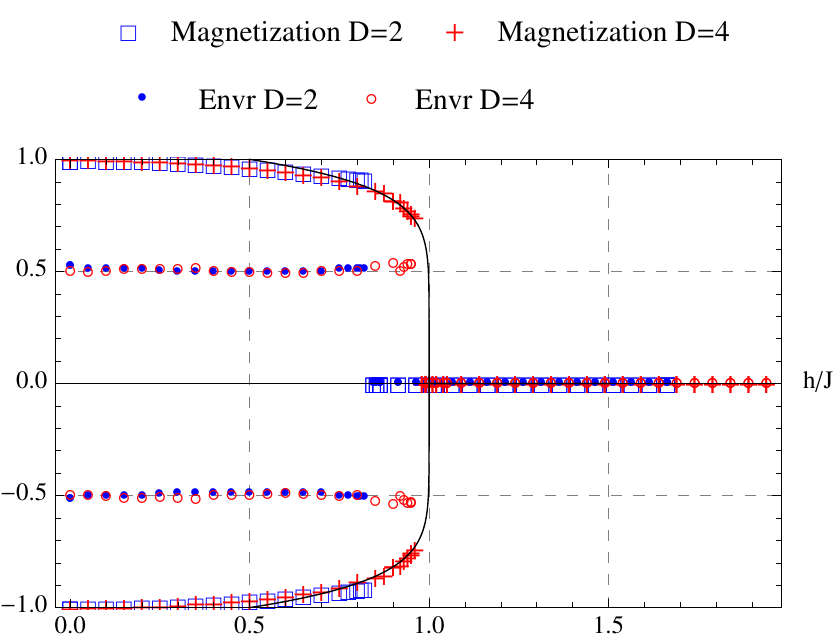}
\caption{\label{1DIsingMag} 
A plot of magnetization and the ``fixed-point'' order paramter
 as a function of $h/J$.
When we increase the internal dimension, we
get a more accurate phase transition point. In both cases, we
get a first-order phase transition. When internal dimension
$D=2$, order parameter is just the off-diagonal term
of the environment matrix (represented by blue dots).
When $D=4$, order parameter 
is the norm of the $2\times 2$ off-diagonal block
in the environment matrix (represented by red circles).
The black line curve is 
$\langle \sigma_i^z \rangle = \pm(2-2h)^{1/8}$.
}
\end{figure}

Here we've also plotted the magnetization as a function of $h/J$ in
Fig. \ref{1DIsingMag}. Note that in the graph, we get a first-order
phase transition for both $D=2$ and $D=4$. This is because
we required our matrices $M$'s in the MPS to have the $\mathbb{Z}_2$
symmetry (recall eqn. \eqref{MPSAltBasis}). This symmetry
requirement favors the symmetric phase, because
symmetry-breaking phase requires $a=c$, so $M$
is block-diagonalized, reducing its effective internal dimension. 
Thus the phase transition point is shifted leftwards, leading to a
first-order transition. As we increase the internal dimension,
we expect the phase transition point to approach $h/J=1$ from
the left.
 
In Fig. \ref{1DIsingMag}, we've also plotted the order parameter
as a function of $h/J$. Note that these ``order parameters" do not
vary as we change $h/J$. This is because the environment matrix
is obtained from enough iterations that it really corresponds to
the fully renormalized wavefunction. The order parameter obtained
from the environment matrix then corresponds actually to the order parameter at the fixed point, thus is always the same until hitting
the phase transition point.

\section{Detecting phases of MPS without knowing the transformation property 
of the matrices}
\label{MPSNoSymm}

In the above, we have assumed that the matrices in the MPS has the symmetry and
studied how to use the symmetry breaking of the environment matrix to detect
the spontaneous symmetry breaking in MPS.  However, in many calculations, such
as the density-matrix-renormalization-group (DMRG) calculation, the resulting
matrices in the MPS do not have the symmetry in the  symmetry breaking phase,
and in general we do not even know how the matrices transform under the
symmetry transformation (since we do not know how the internal indices should
transform under the symmetry).  In this section, we will discuss how to detect
the spontaneous symmetry breaking in MPS, without knowing how the matrices in
the MPS transforms under the symmetry transformation.

\begin{figure}
\includegraphics[scale=0.6]{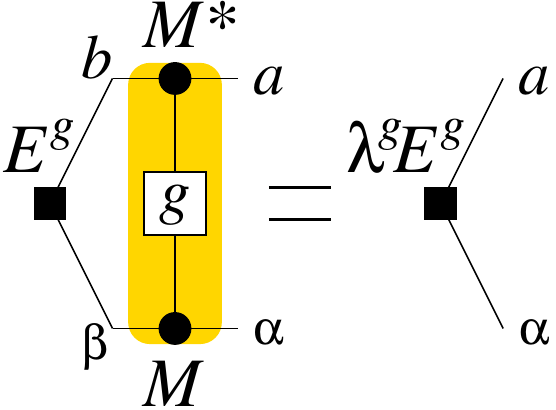}
\caption{\label{1DiterG} A self-consistent condition 
that environment matrix  $E^g$ must satisfy, where
$\lambda^g$ is a scaling factor. 
The part enclosed by the shaded area is the symmetry twisted double-tensor
$T^g_{b\beta,a\alpha}$.}
\end{figure}

Assume we have already obtained the  matrices in the MPS.  We first calculate
the environment matrix $E$ and the scaling factor $\lambda$ using Fig.
\ref{1Diter}.  In general, the  environment matrix $E$ is unique even in the
symmetry breaking state, since the  matrices in the MPS obtained from DMRG in
general already break the symmetry.
Next, we insert the symmetry transformation $g$ (see
(\ref{MtrxSymm})) in the double-tensor to obtain a twisted double-tensor.  The
corresponding twisted environment matrix is denoted as $E^g$ and the twisted
scaling factor as $\lambda^g$ (see Fig. \ref{1DiterG}).

If $|\lambda^g| < |\lambda|$, then the corresponding MPS have a spontaneous
symmetry breaking.  In fact, there is a more direct way to detect symmetry
breaking. Let $P$ and $P^g$ be the matrices defined via the $n^\text{th}$ power
of double-tensor (where $T$ and $T^g$ are viewed as matrices)
\begin{align}
P_{ab,\alpha\beta}=  T^n_{b\beta,a\alpha}, \ \ \ \ \
P^g_{ab,\alpha\beta}=  (T^g)^n_{b\beta,a\alpha}.
\end{align}
If $P$ and $P^g$ have different singular values in large $n$ limit, then the
corresponding MPS break the symmetry explicitly.

If $|\lambda^g| = |\lambda|$, then 
the two environment matrices $E^g$ and $E$ are related by the symmetry
transformation (see (\ref{MtrxSymm}))
\begin{align}
 E=E^g U_g, \ \ \ \ \text{ or } \ \ \ \
 E^{-1}E^g = U_g^\dag.
\end{align}
In fact, we have
\begin{align}
E = U_g^\dag E U_g
\end{align}
If $U_g$ forms a projective representation of the symmetry group $G$, then the
corresponding MPS does not break the symmetry and has a non-trivial SPT order
protected by the on-site symmetry.  If $U_g$ forms a 1D representation of the
symmetry group $G$, then the corresponding MPS does not break the symmetry and
has a non-trivial SPT order protected by translation symmetry (and the on-site
symmetry).

Let us apply the above approach to a MPS state of spin-1 chian, where the
matrix $M^l$, $l=x,y,z$ are given by the Pauli matrices: $M^l=\sigma^l$.
The double-tensor is given by (see Fig. \ref{1Diter})
\begin{align}
 T_{b\beta,a\alpha}=
\sigma^x_{ba}\sigma^x_{\beta\alpha}
-\sigma^y_{ba}\sigma^y_{\beta\alpha}
+\sigma^z_{ba}\sigma^z_{\beta\alpha}
\end{align}
The action of the  double-tensor $ T_{b\beta,a\alpha}$ on the environment
matrix $E_{b\beta}\to T_{b\beta,a\alpha} E_{a\alpha}$
can be written in a matrix form
\begin{align}
 E\to \sum_l \sigma^l E \sigma^l .
\end{align}
We see that $E=2^{-1/2}\sigma^0$ (the 2-by-2 identity matrix)
is the non-degenerate environment matrix with $\lambda=3$.

Now, let us show that
the MPS has a $Z_2^x\times Z_2^z$ symmetry where $Z_2^x$ is generated by $R_x$
-- the $180^\circ$ spin rotation in $S^x$-direction and $Z_2^z$ is generated by
$R_z$ -- the $180^\circ$ spin rotation in $S^z$-direction.
Under the symmetry twists $R_x$ and $R_z$,
the corresponding double-tensors are
\begin{align}
 T^{R_x}_{b\beta,a\alpha}=
-\sigma^x_{ba}\sigma^x_{\beta\alpha}
-\sigma^y_{ba}\sigma^y_{\beta\alpha}
+\sigma^z_{ba}\sigma^z_{\beta\alpha}
\nonumber\\
 T^{R_z}_{b\beta,a\alpha}=
+\sigma^x_{ba}\sigma^x_{\beta\alpha}
-\sigma^y_{ba}\sigma^y_{\beta\alpha}
-\sigma^z_{ba}\sigma^z_{\beta\alpha} .
\end{align}
The corresponding twisted environment matrices are given by
\begin{align}
 E^{R_x}=2^{-1/2} \sigma^x,\ \ \ \
 E^{R_z}=2^{-1/2} \sigma^z.
\end{align}
with $\lambda^{R_x} =\lambda^{R_z}=-3$.  We see that
\begin{align}
 U_{R_x}=\sigma^x, \ \ \ \
 U_{R_z}=\sigma^z,
\end{align}
Since $|\lambda^{R_x}| =|\lambda^{R_z}|=|\lambda|$
and
\begin{align}
 E= U_{R_x}^\dag E U_{R_x}, \ \ \ \
 E= U_{R_z}^\dag E U_{R_z},
\end{align}
we found that the $Z_2^x\times Z_2^z$ symmetry is not broken.  We also see that
$ U_{R_x}, U_{R_z}$ generate a projective representation of $Z_2^x\times
Z_2^z$.  So the MPS is a SPT state protected by $Z_2^x\times Z_2^z$.

\section{A tensor network approach for 1D model}
\label{1DSPT}

In this section, we are going to use an infinite time-evolving block decimation
(iTEBD) approach\cite{V0701} to study 1D models, such as the transverse Ising
model (\ref{1DIsing}). We are going study symmetry breaking by testing if $
|\lambda^g| =|\lambda|$ or $ |\lambda^g| <|\lambda|$.  

\subsection{The iTEBD method}

The iTEBD method is a tensor network version of the DMRG approach.  The
fundamental idea behind the iTEBD method is to use imaginary time evolution to
get the ground state of a two-body Hamiltonian, and to use Singular Value
Decomposition (SVD) to control the internal dimensions.

\begin{figure}
\includegraphics[scale=0.4]{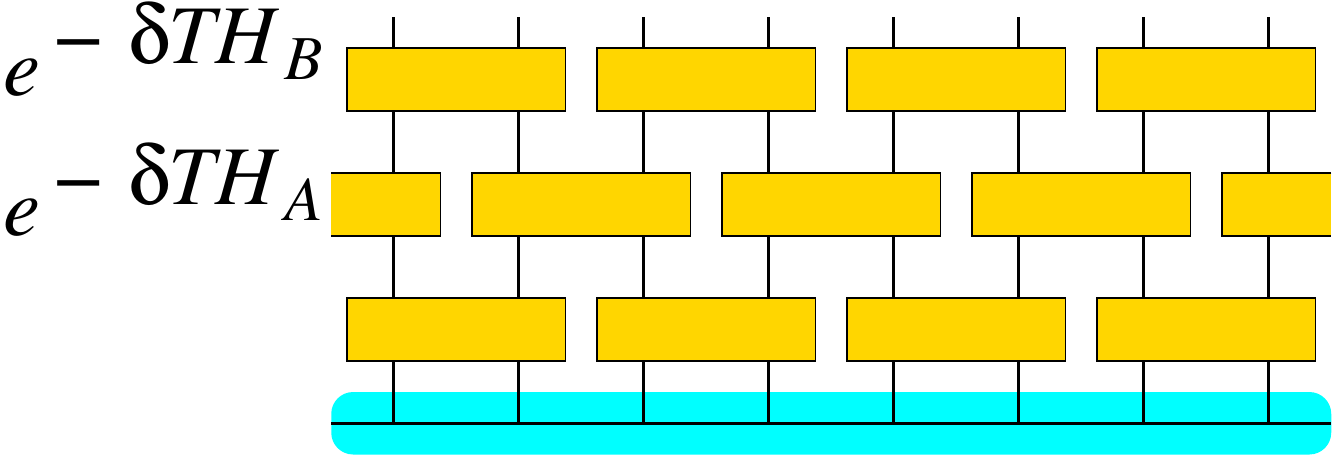}
\caption{\label{HaHb} Applying imaginary time evolution
in a layered structure.}
\end{figure}

Consider any
$1D$ Hamiltonian with only nearest-neigbour interactions,
we can always separate it into two parts, labeled by $H_A$ and $H_B$:
\begin{align}
H &= \sum_i H_{i,i+1} 
= \sum_{i\in odd} H_{i,i+1} + \sum_{i \in even} H_{i,i+1} \nonumber \\
&=H_A + H_B.
\end{align}
This way, either $H_A$ or $H_B$ would have no overlapping terms
within itself.
When the time step $\delta t$ is very tiny, we have:
\begin{align}
e^{-\delta t H} \approx e^{-\delta t H_A} e^{-\delta t H_B} 
\equiv W.
\end{align}
We could then apply imaginary-time evolution layer by layer,
as shown in Fig. \ref{HaHb}. Now within each layer, time-evolution
only operates on non-overlapping neighboring sites. 
Thus the entire problem reduces to a two-site problem. 
 
\begin{figure}
\includegraphics[scale=0.4]{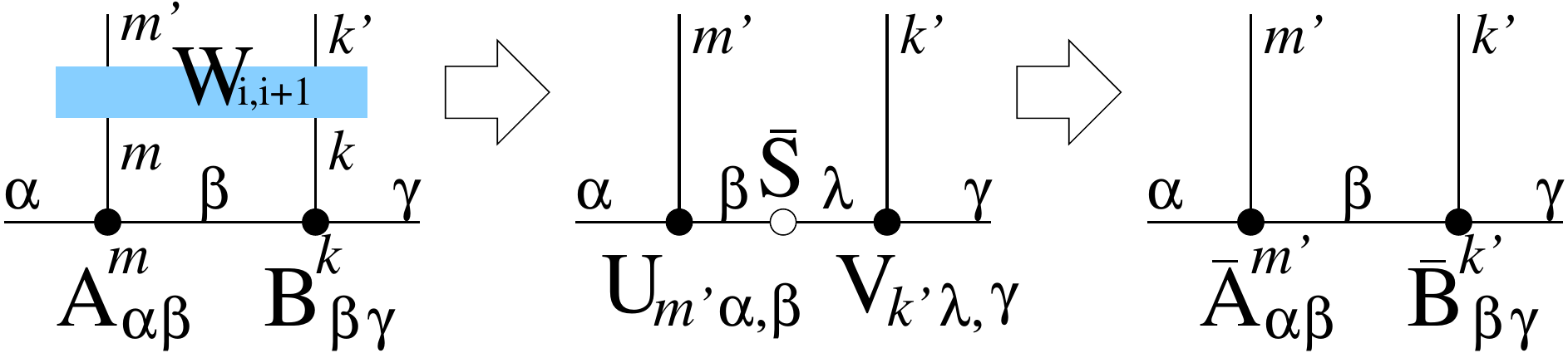}
\caption{\label{DMRG} Time-evolution on two sites. 
Step $1$: apply time-evolution operator.
Step $2$: apply SVD and truncate the singular matrix to
only contain $D$ largest singular values
($\bar S$ denotes the singular matrix after 
truncation).
Step $3$: seperate the singular values into the two sites.}
\end{figure}

The two-site time-evolution is done through Singular Value
Decomposition, see Fig. \ref{DMRG}. We first apply the time-evolution
operator (labeled by $W_{i,i+1}$) on two sites, resulting in a rank-$4$
tensor, $T_{m' \alpha, k' \gamma}$:
\begin{align}
T_{m' \alpha, k' \gamma} = W_{m' k', mk} A^m_{\alpha \beta} B^k_{\beta \gamma}.
\end{align} 
Then we do SVD to split the
rank-$4$ tensor: 
\begin{align}
T_{m' \alpha, k' \gamma} = U_{m' \alpha, \beta} S'_{\beta \lambda} V_{k' \gamma, \lambda}.
\end{align}
Note that after applying SVD, the internal dimension has grown
on the inner link.
We could get back our original internal dimension by keeping only
the $D$ largest singular values of $S'$. We'll call the truncated
matrix $\bar S$.
Finally, we absorb diagnomal matrix $\bar S$  into the two on-site matrices, 
thus completing one step of evolution:
\begin{align}
\bar A^{m'}_{\alpha \beta} = U_{m' \alpha, \beta} \sqrt{\bar S_{\beta}}, \ \  
\bar B^{k'}_{\beta \gamma} = V_{k' \beta, \gamma} \sqrt{\bar S_{\beta}}.
\end{align}
This completes one step in time evolution.

One improvement can be made on the above time-evolution step.
Note that in the truncation process above, we implicitly assumed 
that all bond 
indices are equally important; however, we know that's not the case. 
The ``environment indices'' $\alpha,  \gamma$ do not contribute equally,
and their weights could be naturally included by using the singular values
$S_{\alpha}, S_{\gamma}$ from the previous time-evolution step. 
(This is because in the previous step, 
$\alpha$ and $\gamma$ were inner link indices, and each index naturally
carries a weight according to the previous step of SVD.)

Thus the improved time-evolution step works as follows:
$1.$ First, we scale the ``environment indices'' using singular values 
$S_{\alpha}$ obtained from last step:
\begin{align}
A^m_{\alpha \beta} \rightarrow \sqrt{S_{\alpha}}A^m_{\alpha \beta}, \ \ 
B^m_{\beta \gamma} \rightarrow \sqrt{S_{\gamma}}A^m_{\beta \gamma}.
\end{align}
$2. $ Then apply the
time-evolution step described before.
$3. $ Lastly, we scale the ``enviornment indices'' back, by doing
the following:
\begin{align}
\bar A^m_{\alpha \beta} \rightarrow \sqrt{S^{-1}_{\alpha}} \bar A^m_{\alpha \beta}, \ \ 
\bar B^m_{\beta \gamma} \rightarrow \sqrt{S^{-1}_{\gamma}} \bar B^m_{\beta \gamma}.
\end{align}

\begin{figure}
\includegraphics[scale=0.7]{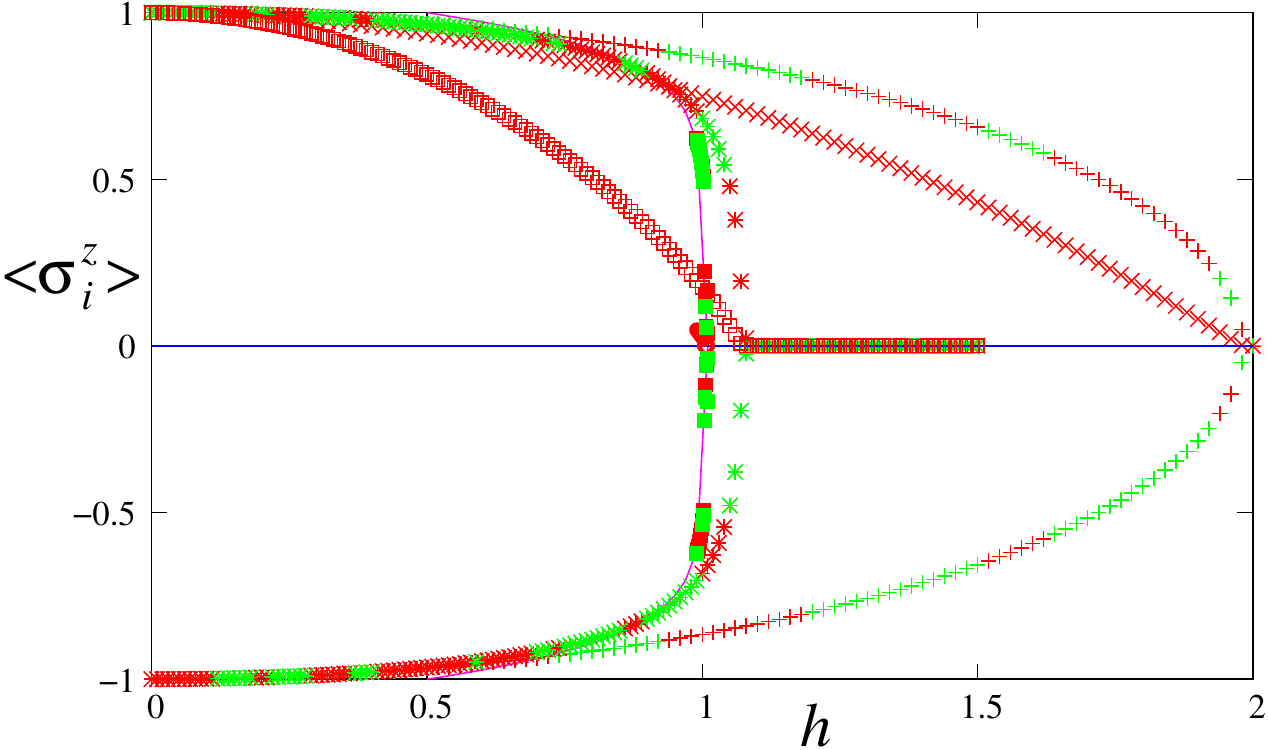}
\caption{\label{tIsing1210} 
The three up-down symmetric curves describe the magnetization of transverse
Ising model $\langle \sigma^z_i\rangle$ as a function of $h$ with $J=1$.  The
``+'' points are for internal dimension $D=1$, ``$+\hskip -2.533mm\times$''
for $D=2$, and filled-box for
 $D=10$.  The other three curves  describe
$\frac{|\lambda|-|\lambda_g|}{|\lambda|}$.  The ``$\times$'' points are for
 $D=1$, ``$\square$'' for $D=2$,
and open-circle near (1,0) are for $D=10$.  The
line curve is $\langle \sigma^z_i\rangle = \pm (2-2h)^{1/8}$.
}
\end{figure}

\begin{figure}
\includegraphics[scale=0.7]{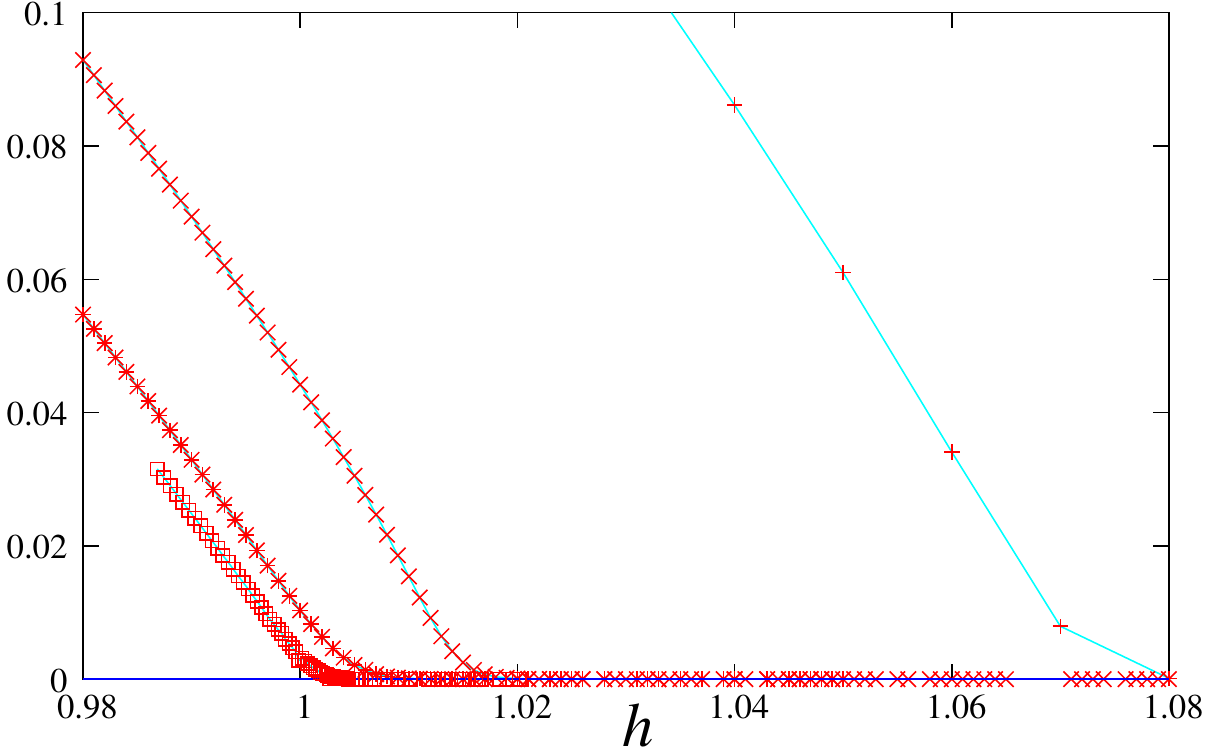}
\caption{\label{tIsing24816la} 
The difference of the scaling factors
$\frac{|\lambda|-|\lambda_g|}{|\lambda|}$ of transverse
Ising model as a function of $h$ with $J=1$.  
The ``+'' points are for $D=2$, 
``$\times$'' for $D=4$, 
``$+\hskip -2.533mm\times$'' for $D=8$,
and ``$\square$'' for $D=16$.
}
\end{figure}
 
\begin{figure}
\includegraphics[scale=0.65]{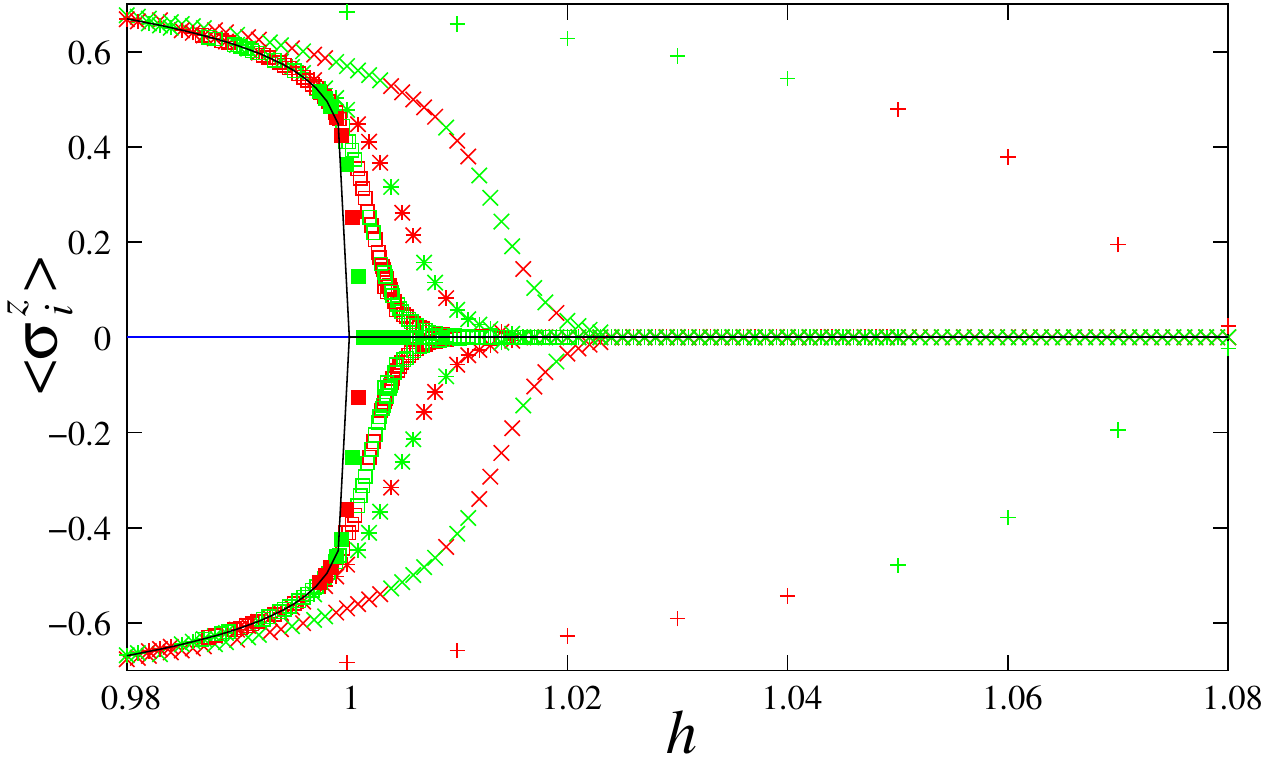}
\caption{\label{tIsing2481632} 
The magnetization of transverse Ising model $\langle \sigma^z_i\rangle$ as a
function of $h$ with $J=1$.  The ``+'' points are for $D=2$, ``$\times$''
for $D=4$, ``$+\hskip -2.533mm\times$'' for $D=8$,
``$\square$'' for $D=16$, 
and ``$\blacksquare$'' for $D=32$. 
The line curve is $\langle \sigma^z_i\rangle = \pm (2-2h)^{1/8}$.
}
\end{figure}

\subsection{The iTEBD results for transverse Ising model}

After applying the imaginary-time evolution steps described above, we will
eventually obtain the tensor that describes the ground state wave function very
well. The next issue is to identify the symmetry breaking order and/or SPT
order in the ground state, using the method discussed before.

For the transverse Ising model, Fig.  \ref{tIsing1210} describes the calculated
$\frac{|\lambda|-|\lambda_g|}{|\lambda|}$ and the magnetization $\langle
\sigma^z_i\rangle$, using the iTEBD approach with various $D$.  Fig.  \ref{tIsing24816la} and Fig.  \ref{tIsing2481632} are
the results near the transition point.  The transition point is found to be
$h_c\approx 1.08$ for $D=2$, $h_c\approx 1.0188$ for $D=4$, $h_c\approx 1.0101$
for $D=8$, 
$h_c\approx 1.0047$ for $D=16$,  
and $h_c\approx 1.0015$ for $D=32$.  
The exact transition point is
at $h_c=1$.  We see that ``order parameter''
$\frac{|\lambda|-|\lambda_g|}{|\lambda|}$ works very well, in identify symmetry
breaking transitions.

\subsection{The iTEBD calculation of 1D model with symmetry-breaking and/or SPT
orders}
 
\begin{figure}
\includegraphics[scale=0.7]{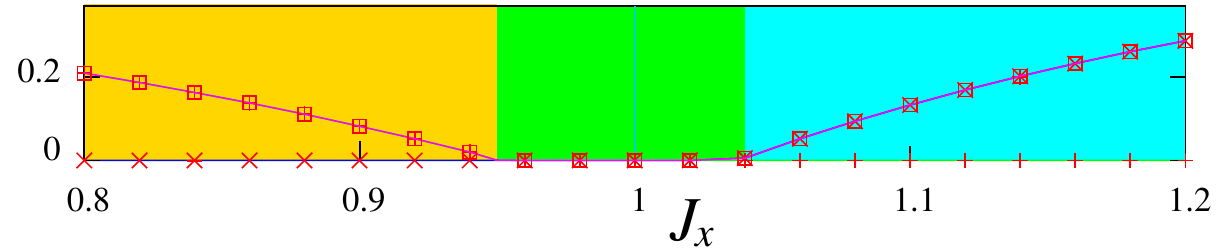}
\caption{\label{Jx8} \label{Jx8} $\frac{|\lambda|-|\lambda_g|}{|\lambda|}$ for the
spin-1 model as a function of $J_x$ with $J_y=J_z=1, J_{zz}=0.4$, calculated
by the iTEBD method with $D=8$.  The ``+'' points are for
symmetry twist $g=R_x$, ``$\times$'' for $g=R_y$, and ``$\square$'' for $g=R_z$.  
When $\frac{|\lambda|-|\lambda_g|}{|\lambda|}=0$, the corresponding symmetry $g$ is not broken.
We see that the phase near $J_x=1$ has the full symmetry.
The phase for smaller $J_x$ has only the $R_y$ symmetry.
The phase for larger $J_x$ has only the $R_x$ symmetry.
}
\end{figure}

\begin{figure}
\includegraphics[scale=0.7]{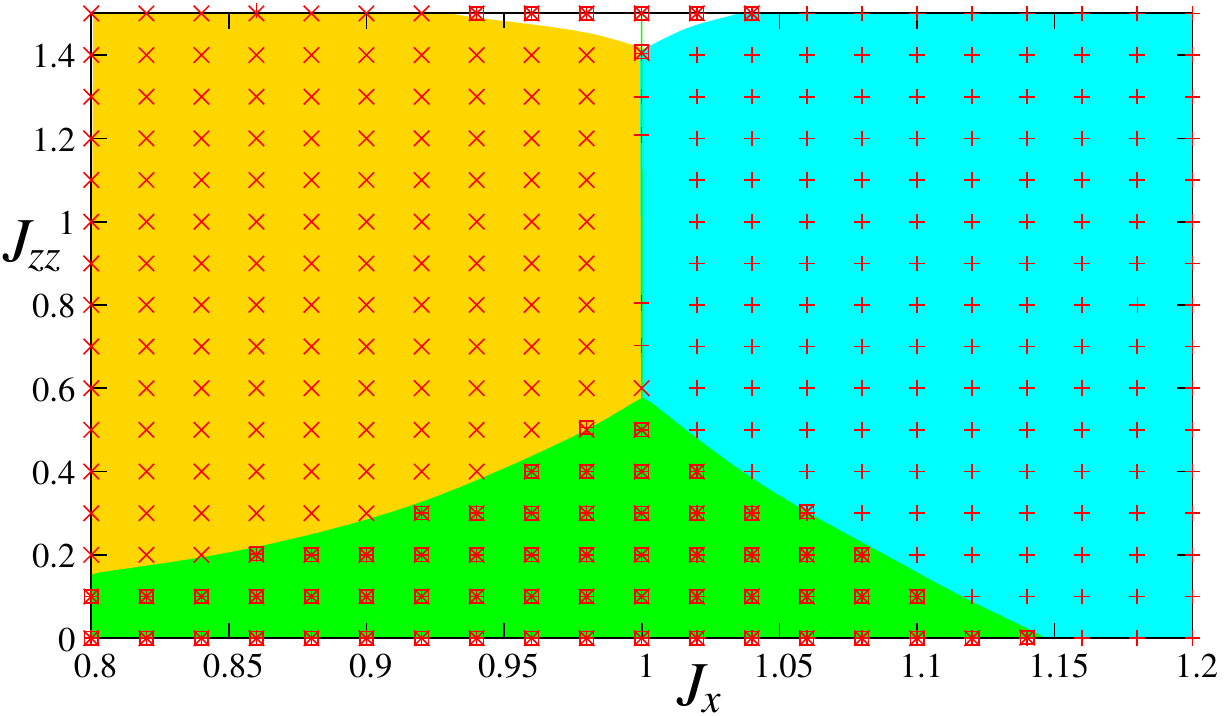}
\caption{\label{JxJzz8} 
The phase diagram for the
spin-1 model in $J_x$-$J_{zz}$ plane, with $J_y=J_z=1$, calculated
by the iTEBD method with $D=8$.  The points mark which
$\frac{|\lambda|-|\lambda_g|}{|\lambda|}=0$.
The ``+'' points are for
symmetry twist $g=R_x$, ``$\times$'' for $g=R_y$, and ``$\square$'' for $g=R_z$.  
The green shaded area and the
white area at the top have the full $R_x,R_y,R_z$ symmetry.
The gold shaded area has only the $R_y$ symmetry, and
the blue shaded area has only the $R_x$ symmetry.
}
\end{figure}

In this section, we are going to use the iTEBD appraoch to study spin-1 model with
$Z_2^x\times Z_2^z$ symmetry:
\begin{align}
\label{spin1}
H=  \sum_{i}
[J_x S_i^x S_{i+1}^x +J_y S_i^y S_{i+1}^y +J_z S_i^z S_{i+1}^z 
+ J_{zz} (S_i^z)^2 ]
\end{align}
The $Z_2^x$ is generated by $180^\circ$ spin-rotation $R_x$ around the $S_x$
axis.  The $Z_2^z$ is generated by $180^\circ$ spin-rotation $R_z$ around the
$S_z$ axis.  The $R_xR_z=R_y$ is the $180^\circ$ spin-rotation around the
$S_y$ axis.

We choose $D=8$ and calculated the tensor $M^m_{\alpha\beta}$ for the ground
state.  To determine if $M^m_{\alpha\beta}$ describes a symmetry breaking state
or not, it is not correct to directly test if $M^m_{\alpha\beta}$ has the
symmetry or not.  This is because even when  $M^m_{\alpha\beta}$ is not
invariant under any symmetry transformation of the form \eqref{MtrxSymm},
$M^m_{\alpha\beta}$ can still describe a symmetric state.

So to determine symmetry of the ground state, we instead calculated
the quantity
$\frac{|\lambda|-|\lambda_g|}{|\lambda|}$ for
symmetry twists $g=R_x,R_y,R_z$ (see Fig. \ref{Jx8}).  We determined the phase
diagram by examine where and which $\frac{|\lambda|-|\lambda_g|}{|\lambda|}$
vanishes. The gold shaded area in Fig. \ref{JxJzz8} only has
the $R_x$ symmetry, since
only $\frac{|\lambda|-|\lambda_{R_x}|}{|\lambda|}=0$.
The blue shaded area in Fig. \ref{JxJzz8} only has
the $R_y$ symmetry since only
$\frac{|\lambda|-|\lambda_{R_y}|}{|\lambda|}=0$.
The green shaded area and the white area  have
$\frac{|\lambda|-|\lambda_g|}{|\lambda|}=0$ for $g=R_x,R_y,R_z$ and have the
full $R_x,R_y,R_z$ symmetry.  In fact the green area is a phase with a
non-trivial SPT order (the Haldane phase).

\section{Application to 2D model with symmetry breaking transition}
\label{Sec2DIsing}

\begin{figure}
\includegraphics[scale=0.8]{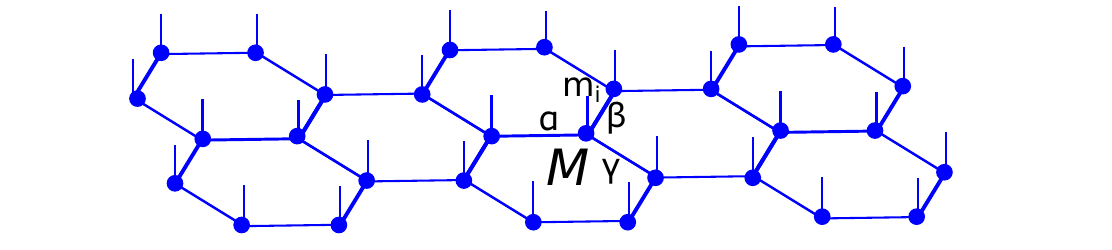}
\caption{\label{TPS} A tensor-network state. 
All the physical sites are represented by dots, physical
degrees of freedom by vertical lines, and internal degrees
of freedom by in-plane links.}
\end{figure}

\begin{figure}
\includegraphics[scale=0.7]{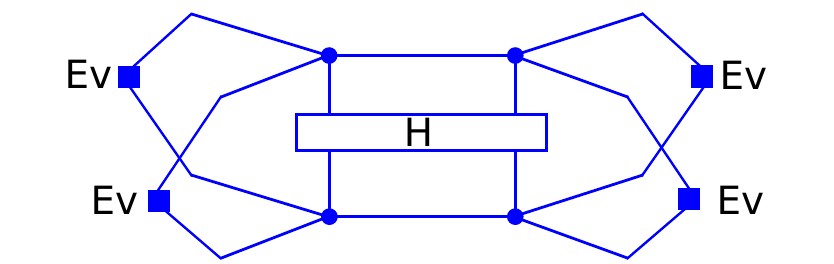}
\caption{\label{2D_energy} The average energy
with total environment tensor $E^{tot}$ in $2D$. 
Here, $E^{tot}$ consists of four environment matrices,
surrounding the two physical sites.}
\end{figure}

Now we want to generalize the above simple picture to $2D$. 
Consider the transverse Ising model
on an infinite honeycomb lattice, with spins living on vertices. 
The Hamiltonian remains the same as equation \eqref{1DIsing}.
Since the ground state of a gapped system in $2D$ could be    
faithfully described by a tensor-network state,
\cite{Tao_DMRG_01, Legeza_DMRG_04, Vidal_AreaLaw_03,
Srednicki_93, Plenio_AreaLaw_05}
for a translation invariant
system, we have (see Figure \ref{TPS}):
\begin{align}
\label{2Dwavefunction}
\Ket{\Psi}=\sum_{ \{m_i\} }\sum_{ \{\alpha, \beta, \gamma\} }
\text{tTr}[\otimes_i M_{\alpha, \beta, \gamma}^{m_i}]  \Ket{ \{m_i\} }
\end{align}
where tensors $M$'s are again labeled by the 
physical degrees of freedoms $m_i$, and $\text{tTr}$ (tensor trace)
contracts over all internal degrees of freedom 
on connected links labeled by $\alpha,\beta$ and $\gamma$. 
Again, we want to do variational calculations with a simple
picture involving
the total environment tensor $E^{tot}$, which now consists of
four environment matrices, see Figure \ref{2D_energy}. 

The key question now is how do we 
obtain a good environment matrix, as we did in $1D$
(recall Figure \ref{1Diter})?
Here we introduce a simple yet powerful iteration process:
assume we have a three-fold
rotational symmetry for tensor $M$, 
then the iteration needs two input matrices,
and gives out only one output, see Figure \ref{2Diter}.
As before, after enough numbers
of iterations, we would reach a final stable ``environment matrix''.

\begin{figure}
\label{2Diter}
\includegraphics[scale=0.7]{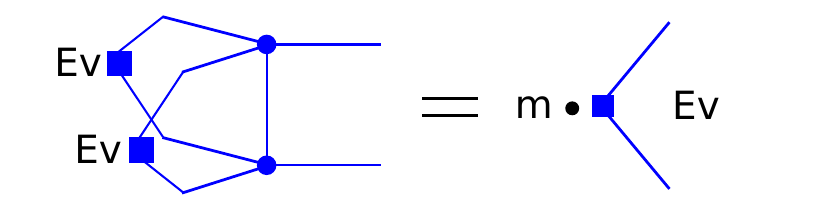}
\caption{\label{2Diter} The self-consistent iteration process for
environment in $2D$. }
\end{figure}

\begin{figure}
\includegraphics[scale=0.6]{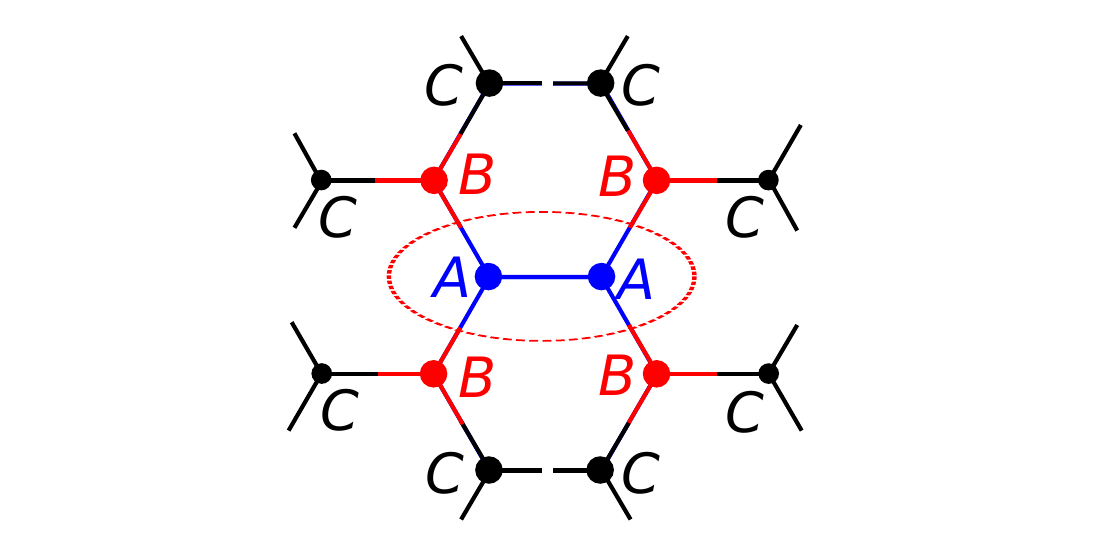}
\caption{\label{Bethe} The iteration process on Bethe lattice.
Note that this is a top-down view, and we have only shown one layer of tensor-network state. Starting from an environment surrouding $C$ tensor, we could iterate to get environment for $B$, and then to $A$.
The circled region is the region of interest.}
\end{figure}

It might be surprising, at first sight, why such a naive 
iteration process would give a reliable environment matrix.
The key however is to realize that this iteration actually gives 
an environment matrix for the infinite Bethe lattice, 
which is a very good first approximation
for our honeycomb lattice (see Figure \ref{Bethe}).
As shown in the graph, the iteration process is actually
equivalent to a self-consistent update for a large cluster
of lattice points, and thus its legitimacy.

Just like in the $1D$ case, here we would also like
to require our tensor-product state to have a 
$\mathbb{Z}_2$ symmetry corresponding to the
spin up-down symmetry.
Similar to the matrix-product state,
on-site symmetry of the ground state also requires tensor
$M$ in the tensor-network state to transform in a special
way:
\begin{align}
\label{TensrSymm}
\sum_{m'}{g_{mm'}M_{\alpha, \beta, \gamma}^{m'}} = 
e^{i\theta_g}\sum_{\alpha',\beta',\gamma'}
{M_{\alpha', \beta', \gamma'}^{m} 
U_g^{\alpha \alpha'} U_g^{\beta \beta'} U_g^{\gamma \gamma'}}.
\end{align}
This is just a tensor generalization of condition \eqref{MtrxSymm}.
As before, $m$ is the physical spin label, $g_{mm'}$ represents
the on-site symmetry and acts in the spin space, $U_g$ 
forms a projective representation of the symmetry group $g$,
and is a unitary matrix acting on internal degrees of freedom labeled
by $\alpha, \beta$ and $\gamma$. 

\begin{figure}
\includegraphics[scale=1.0]{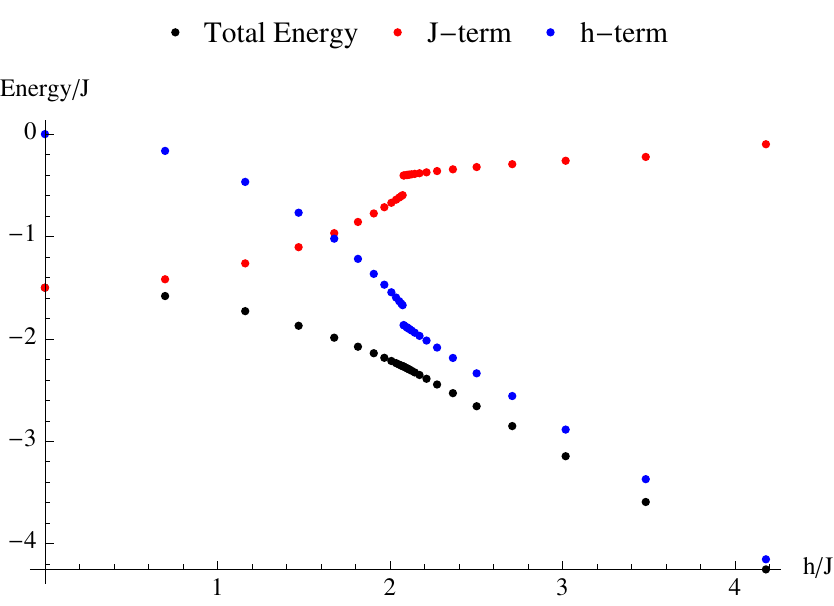}
\caption{\label{2DIsingEnergy} Energy as a function of h/J
for the $2D$ transverse Ising model. 
As in the $1D$ case, h- and J- terms
are individually plotted in the graph as well.
We can see from the graph that our simulation shows
a weak first-order phase transition, with phase
transition point at $h/J=2.09$.}
\end{figure}

For internal dimension of $D=2$,
we can choose $U_g =
\begin{pmatrix}1 & 0 \\ 0 & -1 \end{pmatrix}$.
The most general symmetric tensor
$M^{m_i}_{\alpha, \beta, \gamma}$
satisfying eqn. \eqref{TensrSymm} has $4$
variational parameters and looks like the following:
\begin{align}
M^{\uparrow, \alpha = 1}_{\beta, \gamma} 
&= \begin{pmatrix} a & b \\ b &  c \end{pmatrix}  \ \ 
&&M^{\uparrow, \alpha = 2}_{\beta, \gamma}
= \begin{pmatrix} b & c \\ c &  d \end{pmatrix}, \nonumber \\
M^{\downarrow, \alpha = 1}_{\beta, \gamma} 
&= \begin{pmatrix} a & -b \\ -b &  c \end{pmatrix} \ \ 
&&M^{\uparrow, \alpha = 2}_{\beta, \gamma}
= \begin{pmatrix} -b & c \\ c &  -d \end{pmatrix}.
\end{align}
Here we again assume $M$'s to be symmetric, because
of rotational symmetry.

With the above $M$ tensors,
numerical simulation could again be
run on the $2D$ Ising model.
Following what we did in $1D$, 
we vary $h/J$ in equation \eqref{1DIsing} and minimize
the energy for each value of $h/J$. 
By plotting the two energy terms, a phase diagram could
also be obtained (see Fig. \ref{2DIsingEnergy}).
For internal dimension $D=2$,
the phase transition point occured at $h/J = 2.09$, which
was within $2\%$ error from Quantum Monte Carlo prediction
of $h/J = 2.13$.\cite{Deng_QMC_02} 
The typical runtime on a laptop was just a few seconds.
If we increase the internal dimension to $D=4$ and 
use a completely symmetric tensor with $10$
variational paramters, then we get a phase transition
point at $h/J = 2.12$, within $1\%$ error from
the aforementioned Quantum Monte Carlo calculation.
Here all variational parameters are real.

Following what we did in $D=1$,
here we would also like to comment on 
the symmetry structure of the
environment matrix $E$. 
Recall that $E$ is obtained through
iterations (or self consistent condition) in Fig. \ref{2Diter},
which picks out the $E$ with the largest
absolute value of scaling factor $\lambda$.
One important
difference/simplification in $2D$ is that unlike in $1D$, 
in general, we do not have any degeneracies for $E$
through the iteration equation (Fig. \ref{2Diter}),
since the equation is non-linear. 
Thus in general $E$ obtained is unique, and we do not
need eqn. \eqref{MinEntropy} to fix the basis. 

\begin{figure}
\includegraphics[scale=1.0]{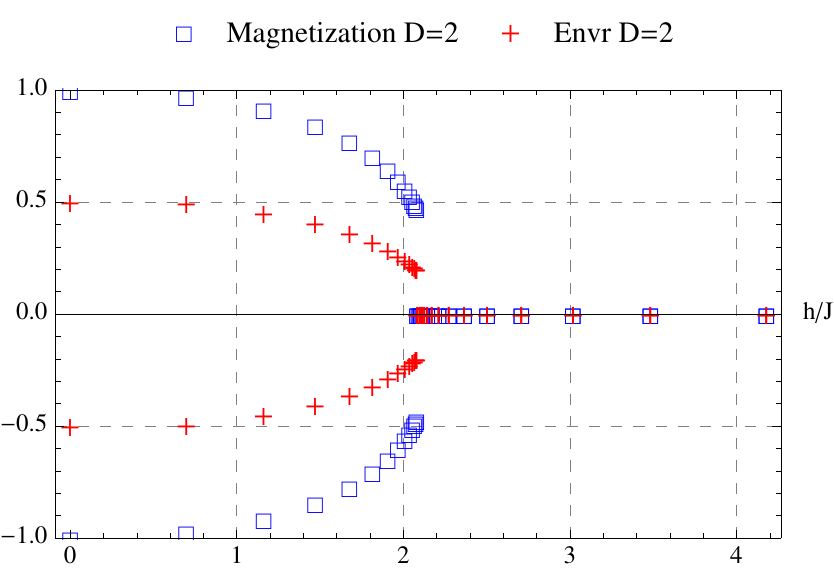}
\caption{\label{2DIsingMag} 
Magnetization and order parameter
as a function of h/J
for the $2D$ transverse Ising model. 
When $D=2$, the order parameter plotted
(represented by red crosses) is
${p/(s+t)}$, which goes to zero
in the symmetric phase.}
\end{figure}

With the above discussion, we can go into the
symmetry structure for our environment matrix $E$
(See Fig. \ref{2DIsingMag}).
For internal dimension being 2, using the iteration
method mentioned above and after energy minimization,
we have in the symmetric phase
$E = \begin{pmatrix} s & 0 \\ 0 &  t \end{pmatrix}$,
which gives an invariant $E$ under eqn. \eqref{EnvrSymm}.
As a result, the total environment tensor is just the direct-product
of them:
\begin{align}
E^{tot} = E \otimes E \otimes E \otimes E.
\end{align}
As for the symmetry breaking phase, 
depending on the initial values of E, we get
either  $E^{g_1} = \begin{pmatrix} \tilde{s} & p \\ p & \tilde{t} \end{pmatrix}$
or  $E^{g_2} = \begin{pmatrix} \tilde{s} & -p \\ -p &  \tilde{t} \end{pmatrix}$,
which transforms into each other under eqn. \eqref{EnvrSymm}.
If we construct the total environment tensor
\begin{align}
E^{tot} = \sum_{g}{E^g \otimes E^g  \otimes E^g \otimes E^g},
\end{align}
then the $\mathbb{Z}_2$ symmetry is restored, but now again the total
environment tensor does not have a pure tensor product form,
as was the case in $1D$.


\section{Application to 2D model with topological order}
\label{Sec2DTop}

We now move on to the non-trivial example of Toric-Code
model in a B-field, with spins
living on links of an infinite $2D$ honeycomb lattice. 
The Hamiltonian is as follows:
\begin{align}
\label{ToricCode}
H= -A \sum_v \prod_{i \in v} \sigma_i^z
-B \sum_p \prod_{j \in p} \sigma_j^x
-h \sum_k \sigma_k^z
\end{align}
where $\sigma$'s are the usual Pauli matrices.
We will first consider the phase diagram by fixing $A \to \infty$
and varying $h/B$ between $[0, \infty]$. 
When $h/B = 0$, we have the original Toric-Code model,
whose ground state is an equal-weight superposition
of all closed loops of down-spins (in the background of up-spins).
When $h/B \to \infty$, we have the spin-polarized state
where all spins are pointing up. 

Later in this
section, we will also consider the case when 
$h/B \to -\infty$ so 
the ground state is the fully packed loop
state, which is an equal weight superposition
of all loop configurations that are fully packed 
(every vertex has a loop passing through).
The question of whether the fully packed loop state
has topological order
or not will then be explored.
 
\begin{figure}
\includegraphics[scale=0.8]{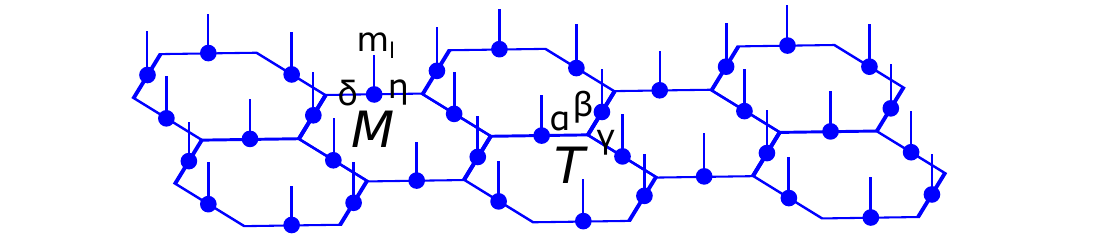}
\caption{\label{TPS_TC} Tensor-product state with spins living
on links. Here physical sites are represented by dots.}
\end{figure}

\begin{figure}
\includegraphics[scale=0.6]{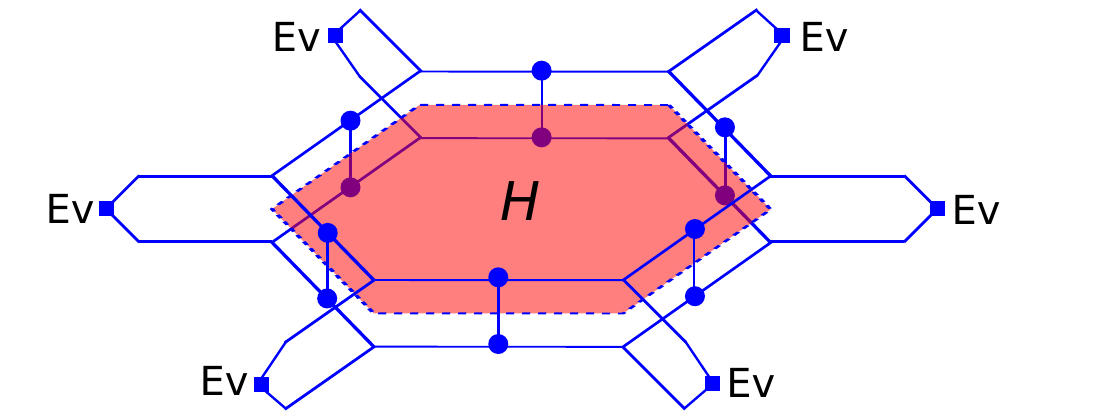}
\caption{\label{TC_energy} Variational energy for Toric-Code
model in a B-field.}
\end{figure}

As in the previous example,
we now try to use a tensor-network state to represent
the ground state of the above Hamiltonian. 
Here since all spins live on links of the lattice,
we will need two tensors $T$ and $M$ to represent
our variational ground state (see Figure \ref{TPS_TC}):
\begin{align}
\label{TCwavefunction}
\Ket{\Psi}=\sum_{ \{m_l\} }
\sum_{ \{\alpha, \beta, \gamma, \delta, \eta\} }
\text{tTr}[\otimes_v T_{\alpha, \beta, \gamma}
\otimes_l M_{\delta, \eta}^{m_l} ] \Ket{ \{m_l\} }
\end{align}
where $v$ labels different vertices, $l$ labels different
links, $\alpha, \beta, \gamma, \delta, \eta$ label
internal degrees of freedom, $m_l$ label physical degrees
of freedom of link $l$, and $\text{tTr}$ contracts over all connected
internal indices.
 Note that due to the $B$ term
in the Hamiltonian \ref{ToricCode}, we will need to include an entire plaquette
in our variational calculation, as shown in Figure \ref{TC_energy}.

Now we start by introducing our tensor ansatz in the simple
case of internal dimension $2$.
In order to enforce the condition that $A \to \infty$ and the
rotational symmetry of the system, we need the following
tensors $T$ and $M$:
\begin{align}
\label{TCansatz}
T_{\alpha, \beta, \gamma} =&
\begin{cases} 1,& if \  \alpha=\beta=\gamma = 0 ; \\
x,& else\ if \  \alpha+\beta+\gamma = 0 \  (\text{mod}\ 2); \\
0,& \text{otherwise}; \end{cases}  \nonumber \\
M_{\delta, \eta}^{\uparrow}=& \begin{pmatrix} 1 & 0 \\ 0 & 0 \end{pmatrix}, \ \ 
M_{\delta, \eta}^{\downarrow}= \begin{pmatrix} 0 & 0 \\ 0 & 1 \end{pmatrix}.
\end{align}
where spin-up and spin-down's are labeled by arrows. 
Note that when $x = 1$, it represents the regular Toric-Code 
ground state\cite{Gu_TPS_09}, whereas when $x = 0$, it represents
the all-spins-up state. 

Before going into our variational calculations, we first note
that our model in equation \eqref{ToricCode} could be mapped
into a transverse Ising model by introducing a new plaquette 
spin operator $\mu_p$, where spins live on the plaquettes 
and $p$ is the plaquette label.\cite{Nayak_TC_07}
By doing the following mapping:
$\prod_{j \in p} \sigma_j^x \to \mu_p^x$,
$\sigma_i^z \to \mu_p^z \mu_{p'}^z$, and consider only the 
$A \to \infty$ sector, our Hamiltonian reduces
to:
\begin{align}
\label{MappingIsing} 
H = -B \sum_p \mu_p^x -h \sum_{<p,p'>} \mu_p^z \mu_{p'}^z,
 \end{align}
which is the familiar transverse Ising model. Note that this
Ising model is now on a $2D$ triangular lattice.

\begin{figure}
\includegraphics[scale=1.0]{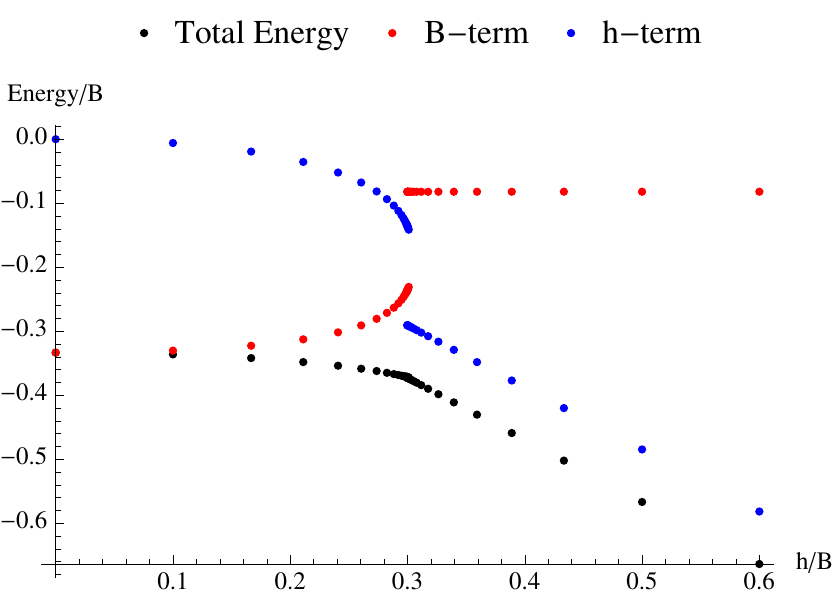}
\caption{\label{TCEnergy} Energy as a function of h/B
for the $2D$ Toric Code model in a magnetic
field. Here we only plotted the region when
$h/B>0$.
We can see from the graph that our simulation shows
a weak first-order phase transition, with phase
transition point at $h/J=0.3$.}
\end{figure}

With the above tensor network ansatz, we could
run our variational scheme on the Toric-Code model.
Just like in the Ising model cases, we vary $h/B$ in
eqn \eqref{ToricCode} (recall that we hold $A \to \infty$)
and minimize the energy for each value of $h/B$.
The environment tensor was calculated in the same way
as before (See Fig. \ref{2Diter}). The only difference here
is that since the Hamiltonian 
\eqref{ToricCode} have
a six-body interaction term, we have to include more sites into
our mean-field calculation(See Figure \ref{TC_energy}).
By plotting the two energy terms as a function of $h/B$,
we get a phase diagram, which is plotted in Fig. \ref{TCEnergy}.
For internal dimension
$D=2$, we got a phase transition point at $B/h=3.33$,
with an error of $30\%$ to the Quantum Monte Carlo
result of $B/h=4.768$.\cite{Deng_QMC_02}
This is not surprising as we only
have one variational parameter.
With internal dimension of $3$ and only two variational
parameters,
our result quickly improved to a phase transition point at
$B/h=4.407$, with an error of less than $8\%$ to the 
Quantum Monte Carlo result.
Note that the Quantum Monte Carlo value
was obtained on the mapped equivalent model (see equation 
\eqref{MappingIsing}) on a $2D$ triangular lattice.

\begin{figure}
\includegraphics[scale=1.0]{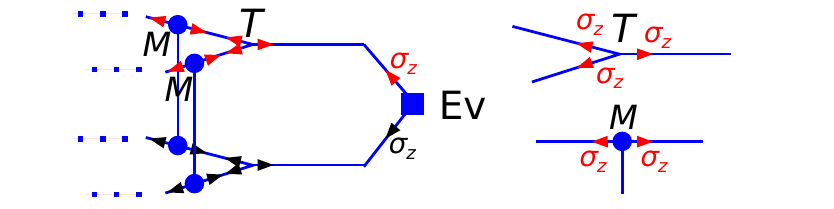}
\caption{\label{SymmTC} 
Internal $\mathbb{Z}_2$ symmetry of the tensors $T$ and $M$
for our Toric-Code model. Here $T$ has a 
$\sigma_z \otimes \sigma_z \otimes \sigma_z$ symmetry,
and $M$ has a $\sigma_z \otimes \sigma_z$ symmetry, both
of which square to identity.
Note that this symmetry transformation 
could independently act
on top and bottom layers of the tensor-network, thus
the environment matrix transforms under
a $\mathbb{Z}_2 \times \mathbb{Z}_2$
group.}
\end{figure}

Now in order to understand the above result better
and to further explore the case when $h/B<0$,
we need to understand the symmetry structure
of both our tensor-product state and
the environment tensor obtained.
Note here that although the ground state doesn't have
a physical $\mathbb{Z}_2$ symmetry, 
the tensor ansatz $T$ and $M$ \eqref{TCansatz}
still need to have an internal $\mathbb{Z}_2$ symmetry,
namely the
``necessary symmetry condition''\cite{Xie_NSymmCond_10} 
(See Figure \ref{SymmTC}):                             
\begin{align}
\label{TCSymm}
T_{\alpha,\beta,\gamma}&=\sum_{\alpha',\beta',\gamma'} 
{T_{\alpha',\beta',\gamma'}\sigma_z^{\alpha \alpha'}
\sigma_z^{\beta \beta'}\sigma_z^{\gamma \gamma'}} \nonumber \\
M_{\delta,\eta}^{m_l}&=\sum_{\delta',\eta'}{M_{\delta',\eta'}
\sigma_z^{\delta \delta'}\sigma_z^{\eta \eta'}}.
\end{align}
Here, the internal  symmetry is represented
by $\sigma_z \otimes \sigma_z \otimes \sigma_z$ for tensor $T$,
and $\sigma_z \otimes \sigma_z$ for tenor $M$,
where $\sigma_z$ is the Pauli matrix.
Since both symmetry actions square to identity,
we refer to the above internal symmetry as a $\mathbb{Z}_2$ symmetry.

The physical reason for tensor ansatz to have the above
``necessary symmetry condition'' is that we want
to make sure local variations of the tensors correspond to
local perturbations of the Hamiltonian.
Tensors that violates the above condition correspond
to non-local perturbation in their Hamiltonian and thus
can not be used to describe physical phase transitions
\cite{Xie_NSymmCond_10}.

It's easy to check that tensors in
equation \eqref{TCansatz} have the above symmetry.
We would then like to ask, with the $T$ and $M$ tensors
satisfying eqn \eqref{TCSymm}, what is the symmetry
structure of the environment matrix? 
Note that unlike the Ising model, here
the internal symmetry of the two layers of our tensor-network
can act independently,  as shown in Figure 
\ref{SymmTC}. Thus the environment matrix
no longer transforms under eqn \eqref{EnvrSymm},
but transforms under a 
$\mathbb{Z}_2 \times \mathbb{Z}_2$ group:
\begin{align}
\label{TCEnvrSymm}
E \to U_g^{\dagger} \cdot E, \ \ or \ \ E \cdot U_g, \ \ 
or \ \  U_g^{\dagger} \cdot E \cdot U_g.
\end{align}
As in the Ising model, we expect 
that in different phases,
the environment matrices $E$'s are either invariant under the 
above transformation,
or undergoes
a permutation.

Our numerical result indeed shows the above feature.
When internal dimension $D=2$, we use the tensor 
ansatz in eqn \eqref{TCansatz} and iteration process
(see Fig. \ref{2Diter}) to get the environment matrix $E$.  
In the confined phase (including spin-polarized state),
we obtain 
$E= \begin{pmatrix} 1 & 0 \\ 0 & 0 \end{pmatrix}$,
which is invariant under eqn. \eqref{TCEnvrSymm}.
As a result, the total environment tensor is
just the direct product of them:
\begin{align}
E^{tot} = E \otimes E \otimes E \otimes E \otimes E \otimes E. \nonumber
\end{align}
In the deconfined phase
(including string-net state), however, we have either
$E^{g_1}= \begin{pmatrix} s & 0 \\ 0 & t \end{pmatrix}$ or
$E^{g_2}= \begin{pmatrix} s & 0 \\ 0 & -t \end{pmatrix}$,
which transforms into each other under eqn. 
\eqref{TCEnvrSymm}. 
We could again construct a total environment tensor
\begin{align}
E^{tot} = \sum_{g}{E^g \otimes E^g  \otimes E^g \otimes E^g \otimes E^g \otimes E^g} \nonumber
\end{align}
that respects the $\mathbb{Z}_2 \times \mathbb{Z}_2$
symmetry, but it does not have a pure tensor product form,
as was the case for Ising model.

In doing the above, we have really constructed a
numerical way to detect topological orders. In the particular
case above, $\mathbb{Z}_2$ topological order is signatured by a
``symmetry breaking'' in the environment matrix,
which breaks the original $\mathbb{Z}_2 \times \mathbb{Z}_2$
symmetry of $E$ (see eqn \eqref{TCEnvrSymm}) down to
$\mathbb{Z}_2$ (see eqn \eqref{EnvrSymm}). 

With this
realization, a natural question to ask is: 
if we now consider negative magnetic field with 
$h/B \to -\infty$,
will the fully
packed loop state has $\mathbb{Z}_2$ topological order?
To answer this question, let us first write down
the ground state wave function of
the fully packed loop state
in tensor form:
\begin{align}
\label{Loopansatz}
T_{\alpha, \beta, \gamma} =&
\begin{cases} 0,& if \  \alpha=\beta=\gamma = 0 ; \\
1,& else\ if \  \alpha+\beta+\gamma = 0 \  (\text{mod}\ 2); \\
0,& \text{otherwise}; \end{cases}  \nonumber \\
M_{\delta, \eta}^{\uparrow}=& \begin{pmatrix} 1 & 0 \\ 0 & 0 \end{pmatrix}, \ \ 
M_{\delta, \eta}^{\downarrow}= \begin{pmatrix} 0 & 0 \\ 0 & 1 \end{pmatrix}.
\end{align}
Note the difference between this and eqn \eqref{TCansatz}:
here we require loops to cover each vertex, 
so $T_{0,0,0}=0$. 

Now to see whether this state 
has $\mathbb{Z}_2$ topological order or not, 
all we need to do
is to calculate its environment matrix through
iteration (see Fig. \ref{2Diter}). Depending
on the initial condition, we obtain either
$E^{g_1}= \begin{pmatrix} 0.38 & 0 \\ 0 & 0.62 \end{pmatrix}$ or
$E^{g_2}= \begin{pmatrix} 0.38 & 0 \\ 0 & -0.62 \end{pmatrix}$,
which again transforms into each other under
eqn \eqref{TCEnvrSymm}.
This means that we are still in the deconfined phase,
and packed loop state has $\mathbb{Z}_2$
topological order.

\begin{figure}
\includegraphics[scale=0.7]{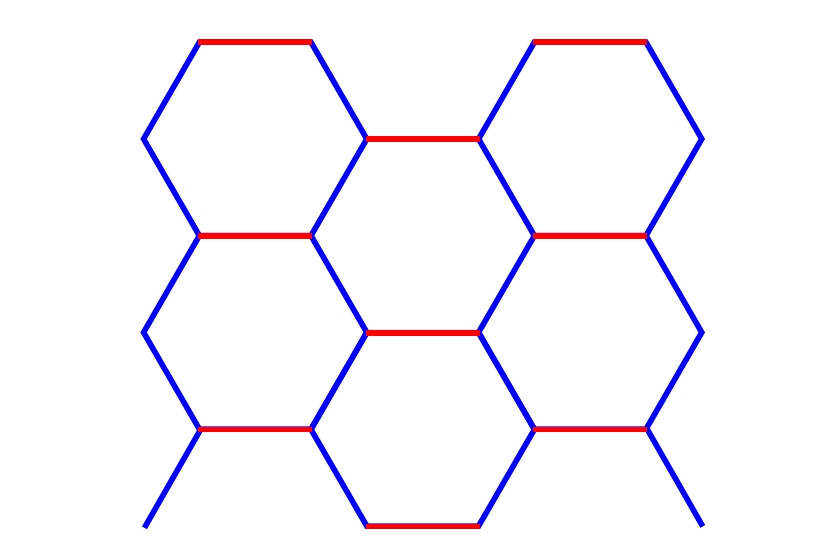}
\caption{\label{StringCrystal} 
The above graph shows a ``string crystal'' state,
where blue lines represent spin-downs forming
vertical strings, and red lines represent spin-ups
forming the background.}
\end{figure}

One may worry that the simple test above
would fail to differentiate the ``string crystal'' state
(see Fig. \ref{StringCrystal}) where string configuration
is stationary,
from the fully packed loop state where
the string configurations are fluctuating. This worry
turns out to be unnecessary through careful
study below. 

Consider a ``string crystal'' state,
with vertical strings formed by
down-spins (shown in Fig. \ref{StringCrystal}). We would
like to study the symmetry structure of the environment
matrix for this state.  
Note that unlike the previous tensor-network
ansatz in eqn \eqref{TCansatz} and \eqref{Loopansatz},
here the tensors no longer have three-fold 
rotational symmetry. 

\begin{figure}
\includegraphics[scale=0.7]{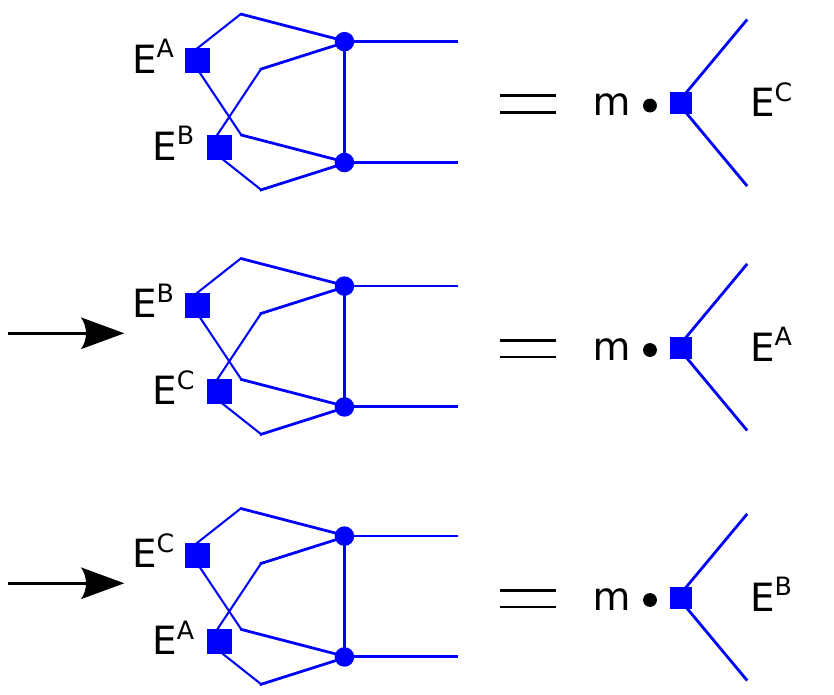}
\caption{\label{2Diter_ABC} The self-consistent iteration process for
environment matrix when the tensors do not have rotational
symmetry. We introduce three different environment matrices,
$E^A$, $E^B$ and $E^C$. One cycle of iteration consists of
three steps: 1. Input $E^A$ and $E^B$ to update $E^C$;
2. Input $E^B$ and $E^C$ to update $E^A$;
3. Input $E^C$ and $E^A$ to update $E^B$.
After iterating for enough number of steps,
all three self-consistent
equation will be simultaneously satisfied.}
\end{figure}

\begin{figure}
\includegraphics[scale=0.6]{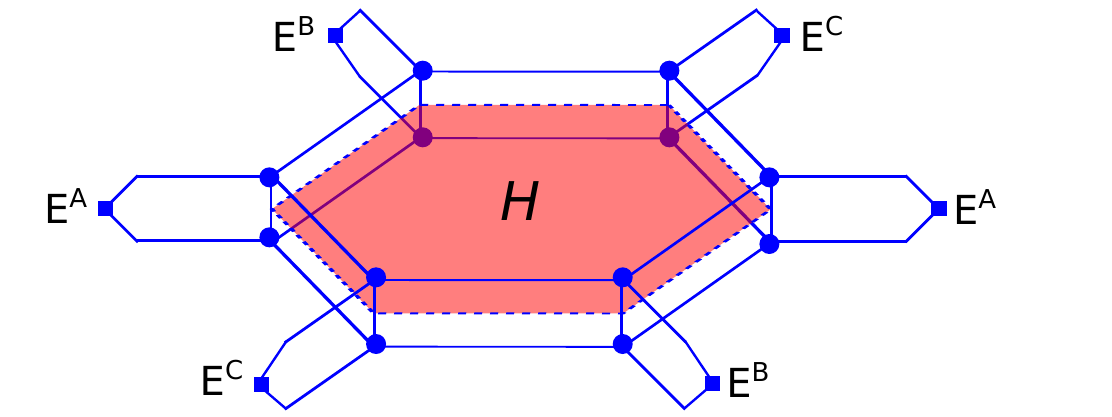}
\caption{\label{TC_energy_ABC} Expectation value
of an operator with non-rotational-symmetric tensor
networks. Note here that there are three different
types of environment matrices.}
\end{figure}

Our method could be easily
generalized to non-rotationally symmetric tensors
by introducing a three-step iteration process, 
shown in Fig. \ref{2Diter_ABC}. (Recall that this is different
from the symmetric iteration process in
Fig. \ref{2Diter}.)
In this three-step iteration process, we introduce
three different environment matrices, which then 
iterate in a cyclic fashion. Now, any physical
quantities could again be calculated by sandwiching
the operator in between two layers of tensor
network states, surrounded by three different
types of environment matrices
shown in Fig. \ref{TC_energy_ABC}, 
and all of our previous analysis
follows.
(Again compare this with Fig. \ref{2D_energy},
where there was only one type of environment matrix.)

With the above three-step iteration process,
the environment matrix of our ``string crystal''
state could then be obtained as follows:
\begin{align}
E^{A}= \begin{pmatrix} 1 & 0 \\ 0 & 0 \end{pmatrix}, \ \ 
E^{B}= E^C= \begin{pmatrix} 0 & 0 \\ 0 & 1 \end{pmatrix}.
\end{align}
Note that the above $E^B$ and $E^C$ do not really
break the $\mathbb{Z}_2 \times \mathbb{Z}_2$
symmetry shown in eqn \eqref{TCEnvrSymm}.
This is because the iteration process for $E^B$
and $E^C$ are both linear, so an overall minus sign
does not affect the iteration result. Thus $E^B$
and $E^C$ could only be determined up to a sign,
which is a fictitious gauge degree of freedom and have
no physical meaning. Thus $E^B \to -E^B$ does not
correspond to breaking the 
$\mathbb{Z}_2 \times \mathbb{Z}_2$ symmetry,
and ``string crystal'' state indeed does not possess
$\mathbb{Z}_2$ topological order.

\section{Conclusions} 

In this paper, we proposed a new signature for
phase transitions between tensor-network states
using the environment matrices. Different
phases are distinctively labeled by different symmetry structures
in the environment matrices. Thus through carefully studying different
symmetry structures of the environment matrix, we could 
identify different phases and
obtain the detailed phase boundaries
for both symmetry-breaking transitions
and topological phase transitions. This greatly helps
us in identifying topological orders or SPT orders from
a generic tensor-network state.

The environment matrix is obtained through
a very simple iteration process using the tensor-network state,
in both $1D$ and  $2D$. 
This iteration process provides a self-consistent environment
matrix that summarizes the contributions from far away sites,
and is like a ``mean-field'' theory for tensor-networks. In the 
same line of thinking, the
environment matrix serves like an ``order parameter''.
What's special about this ``mean-field'' theory is that 
it's suitable for studying long-range entangled states,
and is thus suitable for tackling topological phase transitions.

In $1D$, we demonstrated that this new signature could be easily
combined with existing numerical methods like DMRG or
iTEBD to identify SPT phases.
We first obtain the ground state in a matrix-product form
by applying these $1D$ numerical methods.
Then we calculate the environment matrix,
either through direct iteration process or through 
a twisted iteration process 
(where the symmetry transformation $g$ is sandwiched
in between the double tensor in the iteration process).
By simply comparing the scaling factors in the two
iteration process, we could identify which SPT phase
we are in,
thus providing an easy way to identify SPT orders
directly from a matrix-product state.

In $2D$, the iteration process gives a very efficient
way of calculating variational energies, which in turn
leads to a simple numerical methods in obtaining 
gound state wave function by minimizing the energy.
If we require the ground state tensors to have the proper on-site
symmetry, iteration process could give us environment matrices
that have drastically different symmetry structures, 
labeling different (topological) phases.
Note that the on-site symmetry doesn't have to be a
physical symmetry--- internal gauge symmetry is also valid.

The above numerical method is very general and could be easily
applied to many interesting systems in higher dimension including
$3D$ systems. This will open new doors in numerical study
of higher dimensional systems.



This research is supported by NSF Grant No. DMR-1005541 and
NSFC 11274192.  XGW is also supported by the BMO Financial Group and the John
Templeton Foundation Grant No. 39901.  Research at Perimeter Institute is
supported by the Government of Canada through Industry Canada and by the
Province of Ontario through the Ministry of Research.

\bibliographystyle{apsrev4-1} 
\bibliography{wencross,all,publst,Tensor_Network}

\begin{thebibliography}{49}%
\makeatletter
\providecommand \@ifxundefined [1]{%
 \@ifx{#1\undefined}
}%
\providecommand \@ifnum [1]{%
 \ifnum #1\expandafter \@firstoftwo
 \else \expandafter \@secondoftwo
 \fi
}%
\providecommand \@ifx [1]{%
 \ifx #1\expandafter \@firstoftwo
 \else \expandafter \@secondoftwo
 \fi
}%
\providecommand \natexlab [1]{#1}%
\providecommand \enquote  [1]{``#1''}%
\providecommand \bibnamefont  [1]{#1}%
\providecommand \bibfnamefont [1]{#1}%
\providecommand \citenamefont [1]{#1}%
\providecommand \href@noop [0]{\@secondoftwo}%
\providecommand \href [0]{\begingroup \@sanitize@url \@href}%
\providecommand \@href[1]{\@@startlink{#1}\@@href}%
\providecommand \@@href[1]{\endgroup#1\@@endlink}%
\providecommand \@sanitize@url [0]{\catcode `\\12\catcode `\$12\catcode
  `\&12\catcode `\#12\catcode `\^12\catcode `\_12\catcode `\%12\relax}%
\providecommand \@@startlink[1]{}%
\providecommand \@@endlink[0]{}%
\providecommand \url  [0]{\begingroup\@sanitize@url \@url }%
\providecommand \@url [1]{\endgroup\@href {#1}{\urlprefix }}%
\providecommand \urlprefix  [0]{URL }%
\providecommand \Eprint [0]{\href }%
\providecommand \doibase [0]{http://dx.doi.org/}%
\providecommand \selectlanguage [0]{\@gobble}%
\providecommand \bibinfo  [0]{\@secondoftwo}%
\providecommand \bibfield  [0]{\@secondoftwo}%
\providecommand \translation [1]{[#1]}%
\providecommand \BibitemOpen [0]{}%
\providecommand \bibitemStop [0]{}%
\providecommand \bibitemNoStop [0]{.\EOS\space}%
\providecommand \EOS [0]{\spacefactor3000\relax}%
\providecommand \BibitemShut  [1]{\csname bibitem#1\endcsname}%
\let\auto@bib@innerbib\@empty
\bibitem [{\citenamefont {von Klitzing}\ \emph {et~al.}(1980)\citenamefont {von
  Klitzing}, \citenamefont {Dorda},\ and\ \citenamefont {Pepper}}]{KDP8094}%
  \BibitemOpen
  \bibfield  {author} {\bibinfo {author} {\bibfnamefont {K.}~\bibnamefont {von
  Klitzing}}, \bibinfo {author} {\bibfnamefont {G.}~\bibnamefont {Dorda}}, \
  and\ \bibinfo {author} {\bibfnamefont {M.}~\bibnamefont {Pepper}},\
  }\href@noop {} {\bibfield  {journal} {\bibinfo  {journal} {Phys. Rev. Lett.}\
  }\textbf {\bibinfo {volume} {45}},\ \bibinfo {pages} {494} (\bibinfo {year}
  {1980})}\BibitemShut {NoStop}%
\bibitem [{\citenamefont {Tsui}\ \emph {et~al.}(1982)\citenamefont {Tsui},
  \citenamefont {Stormer},\ and\ \citenamefont {Gossard}}]{TSG8259}%
  \BibitemOpen
  \bibfield  {author} {\bibinfo {author} {\bibfnamefont {D.~C.}\ \bibnamefont
  {Tsui}}, \bibinfo {author} {\bibfnamefont {H.~L.}\ \bibnamefont {Stormer}}, \
  and\ \bibinfo {author} {\bibfnamefont {A.~C.}\ \bibnamefont {Gossard}},\
  }\href@noop {} {\bibfield  {journal} {\bibinfo  {journal} {Phys. Rev. Lett.}\
  }\textbf {\bibinfo {volume} {48}},\ \bibinfo {pages} {1559} (\bibinfo {year}
  {1982})}\BibitemShut {NoStop}%
\bibitem [{\citenamefont {Kane}\ and\ \citenamefont {Mele}(2005)}]{KM0502}%
  \BibitemOpen
  \bibfield  {author} {\bibinfo {author} {\bibfnamefont {C.~L.}\ \bibnamefont
  {Kane}}\ and\ \bibinfo {author} {\bibfnamefont {E.~J.}\ \bibnamefont
  {Mele}},\ }\href@noop {} {\bibfield  {journal} {\bibinfo  {journal} {Phys.
  Rev. Lett.}\ }\textbf {\bibinfo {volume} {95}},\ \bibinfo {pages} {146802}
  (\bibinfo {year} {2005})},\ \Eprint {http://arxiv.org/abs/cond-mat/0506581}
  {cond-mat/0506581} \BibitemShut {NoStop}%
\bibitem [{\citenamefont {Bernevig}\ and\ \citenamefont
  {Zhang}(2006)}]{BZ0602}%
  \BibitemOpen
  \bibfield  {author} {\bibinfo {author} {\bibfnamefont {B.~A.}\ \bibnamefont
  {Bernevig}}\ and\ \bibinfo {author} {\bibfnamefont {S.-C.}\ \bibnamefont
  {Zhang}},\ }\href@noop {} {\bibfield  {journal} {\bibinfo  {journal} {Phys.
  Rev. Lett.}\ }\textbf {\bibinfo {volume} {96}},\ \bibinfo {pages} {106802}
  (\bibinfo {year} {2006})},\ \Eprint {http://arxiv.org/abs/cond-mat/0504147}
  {cond-mat/0504147} \BibitemShut {NoStop}%
\bibitem [{\citenamefont {Moore}\ and\ \citenamefont {Balents}(2007)}]{MB0706}%
  \BibitemOpen
  \bibfield  {author} {\bibinfo {author} {\bibfnamefont {J.~E.}\ \bibnamefont
  {Moore}}\ and\ \bibinfo {author} {\bibfnamefont {L.}~\bibnamefont
  {Balents}},\ }\href@noop {} {\bibfield  {journal} {\bibinfo  {journal} {Phys.
  Rev. B}\ }\textbf {\bibinfo {volume} {75}},\ \bibinfo {pages} {121306}
  (\bibinfo {year} {2007})},\ \Eprint {http://arxiv.org/abs/cond-mat/0607314}
  {cond-mat/0607314} \BibitemShut {NoStop}%
\bibitem [{\citenamefont {Fu}\ \emph {et~al.}(2007)\citenamefont {Fu},
  \citenamefont {Kane},\ and\ \citenamefont {Mele}}]{FKM0703}%
  \BibitemOpen
  \bibfield  {author} {\bibinfo {author} {\bibfnamefont {L.}~\bibnamefont
  {Fu}}, \bibinfo {author} {\bibfnamefont {C.~L.}\ \bibnamefont {Kane}}, \ and\
  \bibinfo {author} {\bibfnamefont {E.~J.}\ \bibnamefont {Mele}},\ }\href@noop
  {} {\bibfield  {journal} {\bibinfo  {journal} {Phys. Rev. Lett.}\ }\textbf
  {\bibinfo {volume} {98}},\ \bibinfo {pages} {106803} (\bibinfo {year}
  {2007})},\ \Eprint {http://arxiv.org/abs/cond-mat/0607699} {cond-mat/0607699}
  \BibitemShut {NoStop}%
\bibitem [{\citenamefont {Kitaev}(2009)}]{K0986}%
  \BibitemOpen
  \bibfield  {author} {\bibinfo {author} {\bibfnamefont {A.}~\bibnamefont
  {Kitaev}},\ }in\ \href@noop {} {\emph {\bibinfo {booktitle} {Advances in
  Theoretical Physics: Landau Memorial Conference, Chernogolovka, Russia,
  2008}}},\ Vol.\ \bibinfo {volume} {AIP Conf. Proc. No. 1134},\ \bibinfo
  {editor} {edited by\ \bibinfo {editor} {\bibfnamefont {V.}~\bibnamefont
  {Lebedev}}\ and\ \bibinfo {editor} {\bibfnamefont {M.}~\bibnamefont
  {Feigel’man}}}\ (\bibinfo  {publisher} {AIP},\ \bibinfo {address}
  {Melville, NY},\ \bibinfo {year} {2009})\ p.~\bibinfo {pages} {22},\ \Eprint
  {http://arxiv.org/abs/arXiv:0901.2686} {arXiv:0901.2686} \BibitemShut
  {NoStop}%
\bibitem [{\citenamefont {Ryu}\ \emph {et~al.}(2009)\citenamefont {Ryu},
  \citenamefont {Schnyder}, \citenamefont {Furusaki},\ and\ \citenamefont
  {Ludwig}}]{RSF0957}%
  \BibitemOpen
  \bibfield  {author} {\bibinfo {author} {\bibfnamefont {S.}~\bibnamefont
  {Ryu}}, \bibinfo {author} {\bibfnamefont {A.}~\bibnamefont {Schnyder}},
  \bibinfo {author} {\bibfnamefont {A.}~\bibnamefont {Furusaki}}, \ and\
  \bibinfo {author} {\bibfnamefont {A.}~\bibnamefont {Ludwig}},\ }\href
  {\doibase 10.1088/1367-2630/12/6/065010} {\bibfield  {journal} {\bibinfo
  {journal} {New J. Phys.}\ }\textbf {\bibinfo {volume} {12}},\ \bibinfo
  {pages} {065010} (\bibinfo {year} {2009})},\ \Eprint
  {http://arxiv.org/abs/arXiv:0912.2157} {arXiv:0912.2157} \BibitemShut
  {NoStop}%
\bibitem [{\citenamefont {Wen}(1989)}]{W8987}%
  \BibitemOpen
  \bibfield  {author} {\bibinfo {author} {\bibfnamefont {X.-G.}\ \bibnamefont
  {Wen}},\ }\href@noop {} {\bibfield  {journal} {\bibinfo  {journal} {Phys.
  Rev. B}\ }\textbf {\bibinfo {volume} {40}},\ \bibinfo {pages} {7387}
  (\bibinfo {year} {1989})}\BibitemShut {NoStop}%
\bibitem [{\citenamefont {Wen}\ and\ \citenamefont {Niu}(1990)}]{WN9077}%
  \BibitemOpen
  \bibfield  {author} {\bibinfo {author} {\bibfnamefont {X.-G.}\ \bibnamefont
  {Wen}}\ and\ \bibinfo {author} {\bibfnamefont {Q.}~\bibnamefont {Niu}},\
  }\href@noop {} {\bibfield  {journal} {\bibinfo  {journal} {Phys. Rev. B}\
  }\textbf {\bibinfo {volume} {41}},\ \bibinfo {pages} {9377} (\bibinfo {year}
  {1990})}\BibitemShut {NoStop}%
\bibitem [{\citenamefont {Wen}(1990)}]{W9039}%
  \BibitemOpen
  \bibfield  {author} {\bibinfo {author} {\bibfnamefont {X.-G.}\ \bibnamefont
  {Wen}},\ }\href@noop {} {\bibfield  {journal} {\bibinfo  {journal} {Int. J.
  Mod. Phys. B}\ }\textbf {\bibinfo {volume} {4}},\ \bibinfo {pages} {239}
  (\bibinfo {year} {1990})}\BibitemShut {NoStop}%
\bibitem [{\citenamefont {Keski-Vakkuri}\ and\ \citenamefont
  {Wen}(1993)}]{KW9327}%
  \BibitemOpen
  \bibfield  {author} {\bibinfo {author} {\bibfnamefont {E.}~\bibnamefont
  {Keski-Vakkuri}}\ and\ \bibinfo {author} {\bibfnamefont {X.-G.}\ \bibnamefont
  {Wen}},\ }\href@noop {} {\bibfield  {journal} {\bibinfo  {journal} {Int. J.
  Mod. Phys. B}\ }\textbf {\bibinfo {volume} {7}},\ \bibinfo {pages} {4227}
  (\bibinfo {year} {1993})}\BibitemShut {NoStop}%
\bibitem [{\citenamefont {Levin}\ and\ \citenamefont
  {Wen}(2005)}]{Levin_Strnt_05}%
  \BibitemOpen
  \bibfield  {author} {\bibinfo {author} {\bibfnamefont {M.~A.}\ \bibnamefont
  {Levin}}\ and\ \bibinfo {author} {\bibfnamefont {X.-G.}\ \bibnamefont
  {Wen}},\ }\href@noop {} {\bibfield  {journal} {\bibinfo  {journal} {Phys.
  Rev. B}\ }\textbf {\bibinfo {volume} {71}},\ \bibinfo {pages} {045110}
  (\bibinfo {year} {2005})}\BibitemShut {NoStop}%
\bibitem [{\citenamefont {Chen}\ \emph
  {et~al.}(2010{\natexlab{a}})\citenamefont {Chen}, \citenamefont {Gu},\ and\
  \citenamefont {Wen}}]{Xie_LU_10}%
  \BibitemOpen
  \bibfield  {author} {\bibinfo {author} {\bibfnamefont {X.}~\bibnamefont
  {Chen}}, \bibinfo {author} {\bibfnamefont {Z.-C.}\ \bibnamefont {Gu}}, \ and\
  \bibinfo {author} {\bibfnamefont {X.-G.}\ \bibnamefont {Wen}},\ }\href@noop
  {} {\bibfield  {journal} {\bibinfo  {journal} {Phys. Rev. B}\ }\textbf
  {\bibinfo {volume} {82}},\ \bibinfo {pages} {155138} (\bibinfo {year}
  {2010}{\natexlab{a}})}\BibitemShut {NoStop}%
\bibitem [{\citenamefont {Liu}\ \emph {et~al.}()\citenamefont {Liu},
  \citenamefont {Wang}, \citenamefont {You},\ and\ \citenamefont
  {Wen}}]{Liu_ModTrans_13}%
  \BibitemOpen
  \bibfield  {author} {\bibinfo {author} {\bibfnamefont {F.}~\bibnamefont
  {Liu}}, \bibinfo {author} {\bibfnamefont {Z.}~\bibnamefont {Wang}}, \bibinfo
  {author} {\bibfnamefont {Y.-Z.}\ \bibnamefont {You}}, \ and\ \bibinfo
  {author} {\bibfnamefont {X.-G.}\ \bibnamefont {Wen}},\ }\href@noop {}
  {\bibinfo  {journal} {arXiv:1303.0829}\ }\BibitemShut {NoStop}%
\bibitem [{\citenamefont {Gu}\ \emph {et~al.}()\citenamefont {Gu},
  \citenamefont {Wang},\ and\ \citenamefont {Wen}}]{Gu_FermiLU_10}%
  \BibitemOpen
\bibfield  {journal} {  }\bibfield  {author} {\bibinfo {author} {\bibfnamefont
  {Z.-C.}\ \bibnamefont {Gu}}, \bibinfo {author} {\bibfnamefont
  {Z.}~\bibnamefont {Wang}}, \ and\ \bibinfo {author} {\bibfnamefont {X.-G.}\
  \bibnamefont {Wen}},\ }\href@noop {} {\bibinfo  {journal} {arXiv:1010.1517}\
  }\BibitemShut {NoStop}%
\bibitem [{\citenamefont {Kong}\ and\ \citenamefont {Wen}(2014)}]{KW1458}%
  \BibitemOpen
\bibfield  {journal} {  }\bibfield  {author} {\bibinfo {author} {\bibfnamefont
  {L.}~\bibnamefont {Kong}}\ and\ \bibinfo {author} {\bibfnamefont {X.-G.}\
  \bibnamefont {Wen}},\ }\href@noop {} {\  (\bibinfo {year} {2014})},\ \Eprint
  {http://arxiv.org/abs/arXiv:1405.5858} {arXiv:1405.5858} \BibitemShut
  {NoStop}%
\bibitem [{\citenamefont {{Barkeshli}}\ \emph {et~al.}(2014)\citenamefont
  {{Barkeshli}}, \citenamefont {{Bonderson}}, \citenamefont {{Cheng}},\ and\
  \citenamefont {{Wang}}}]{BBC1440}%
  \BibitemOpen
  \bibfield  {author} {\bibinfo {author} {\bibfnamefont {M.}~\bibnamefont
  {{Barkeshli}}}, \bibinfo {author} {\bibfnamefont {P.}~\bibnamefont
  {{Bonderson}}}, \bibinfo {author} {\bibfnamefont {M.}~\bibnamefont
  {{Cheng}}}, \ and\ \bibinfo {author} {\bibfnamefont {Z.}~\bibnamefont
  {{Wang}}},\ }\href@noop {} {\  (\bibinfo {year} {2014})},\ \Eprint
  {http://arxiv.org/abs/1410.4540} {arXiv:1410.4540} \BibitemShut {NoStop}%
\bibitem [{\citenamefont {Chen}\ \emph {et~al.}(2012)\citenamefont {Chen},
  \citenamefont {Gu}, \citenamefont {Liu},\ and\ \citenamefont
  {Wen}}]{Xie_SPT_12}%
  \BibitemOpen
  \bibfield  {author} {\bibinfo {author} {\bibfnamefont {X.}~\bibnamefont
  {Chen}}, \bibinfo {author} {\bibfnamefont {Z.-C.}\ \bibnamefont {Gu}},
  \bibinfo {author} {\bibfnamefont {Z.-X.}\ \bibnamefont {Liu}}, \ and\
  \bibinfo {author} {\bibfnamefont {X.-G.}\ \bibnamefont {Wen}},\ }\href@noop
  {} {\bibfield  {journal} {\bibinfo  {journal} {Science}\ }\textbf {\bibinfo
  {volume} {338}},\ \bibinfo {pages} {1604} (\bibinfo {year}
  {2012})}\BibitemShut {NoStop}%
\bibitem [{\citenamefont {Kapustin}(2014)}]{K1459}%
  \BibitemOpen
  \bibfield  {author} {\bibinfo {author} {\bibfnamefont {A.}~\bibnamefont
  {Kapustin}},\ }\href@noop {} {\  (\bibinfo {year} {2014})},\ \Eprint
  {http://arxiv.org/abs/arXiv:1404.6659} {arXiv:1404.6659} \BibitemShut
  {NoStop}%
\bibitem [{\citenamefont {Wen}(2014{\natexlab{a}})}]{W1477}%
  \BibitemOpen
  \bibfield  {author} {\bibinfo {author} {\bibfnamefont {X.-G.}\ \bibnamefont
  {Wen}},\ }\href@noop {} {\  (\bibinfo {year} {2014}{\natexlab{a}})},\ \Eprint
  {http://arxiv.org/abs/arXiv:1410.8477} {arXiv:1410.8477} \BibitemShut
  {NoStop}%
\bibitem [{\citenamefont {Gu}\ and\ \citenamefont
  {Wen}(2009)}]{GU_CnrDoubleLine_09}%
  \BibitemOpen
  \bibfield  {author} {\bibinfo {author} {\bibfnamefont {Z.-C.}\ \bibnamefont
  {Gu}}\ and\ \bibinfo {author} {\bibfnamefont {X.-G.}\ \bibnamefont {Wen}},\
  }\href@noop {} {\bibfield  {journal} {\bibinfo  {journal} {Phys. Rev. B}\
  }\textbf {\bibinfo {volume} {80}},\ \bibinfo {pages} {155131} (\bibinfo
  {year} {2009})}\BibitemShut {NoStop}%
\bibitem [{\citenamefont {Wen}(2012)}]{W1221}%
  \BibitemOpen
  \bibfield  {author} {\bibinfo {author} {\bibfnamefont {X.-G.}\ \bibnamefont
  {Wen}},\ }\href@noop {} {\  (\bibinfo {year} {2012})},\ \Eprint
  {http://arxiv.org/abs/arXiv:1212.5121} {arXiv:1212.5121} \BibitemShut
  {NoStop}%
\bibitem [{\citenamefont {Zhang}\ \emph {et~al.}(2012)\citenamefont {Zhang},
  \citenamefont {Grover}, \citenamefont {Turner}, \citenamefont {Oshikawa},\
  and\ \citenamefont {Vishwanath}}]{ZGT1251}%
  \BibitemOpen
  \bibfield  {author} {\bibinfo {author} {\bibfnamefont {Y.}~\bibnamefont
  {Zhang}}, \bibinfo {author} {\bibfnamefont {T.}~\bibnamefont {Grover}},
  \bibinfo {author} {\bibfnamefont {A.}~\bibnamefont {Turner}}, \bibinfo
  {author} {\bibfnamefont {M.}~\bibnamefont {Oshikawa}}, \ and\ \bibinfo
  {author} {\bibfnamefont {A.}~\bibnamefont {Vishwanath}},\ }\href {\doibase
  10.1103/PhysRevB.85.235151} {\bibfield  {journal} {\bibinfo  {journal} {Phys.
  Rev. B}\ }\textbf {\bibinfo {volume} {85}},\ \bibinfo {pages} {235151}
  (\bibinfo {year} {2012})},\ \Eprint {http://arxiv.org/abs/arXiv:1111.2342}
  {arXiv:1111.2342} \BibitemShut {NoStop}%
\bibitem [{\citenamefont {Cincio}\ and\ \citenamefont {Vidal}(2013)}]{CV1223}%
  \BibitemOpen
  \bibfield  {author} {\bibinfo {author} {\bibfnamefont {L.}~\bibnamefont
  {Cincio}}\ and\ \bibinfo {author} {\bibfnamefont {G.}~\bibnamefont {Vidal}},\
  }\href@noop {} {\bibfield  {journal} {\bibinfo  {journal} {Phys. Rev. Lett.}\
  }\textbf {\bibinfo {volume} {110}},\ \bibinfo {pages} {067208} (\bibinfo
  {year} {2013})},\ \Eprint {http://arxiv.org/abs/arXiv:1208.2623}
  {arXiv:1208.2623} \BibitemShut {NoStop}%
\bibitem [{\citenamefont {Zaletel}\ \emph {et~al.}(2012)\citenamefont
  {Zaletel}, \citenamefont {Mong},\ and\ \citenamefont {Pollmann}}]{ZMP1233}%
  \BibitemOpen
  \bibfield  {author} {\bibinfo {author} {\bibfnamefont {M.~P.}\ \bibnamefont
  {Zaletel}}, \bibinfo {author} {\bibfnamefont {R.~S.~K.}\ \bibnamefont
  {Mong}}, \ and\ \bibinfo {author} {\bibfnamefont {F.}~\bibnamefont
  {Pollmann}},\ }\href@noop {} {\  (\bibinfo {year} {2012})},\ \Eprint
  {http://arxiv.org/abs/arXiv:1211.3733} {arXiv:1211.3733} \BibitemShut
  {NoStop}%
\bibitem [{\citenamefont {Tu}\ \emph {et~al.}(2013)\citenamefont {Tu},
  \citenamefont {Zhang},\ and\ \citenamefont {Qi}}]{TZQ1251}%
  \BibitemOpen
  \bibfield  {author} {\bibinfo {author} {\bibfnamefont {H.-H.}\ \bibnamefont
  {Tu}}, \bibinfo {author} {\bibfnamefont {Y.}~\bibnamefont {Zhang}}, \ and\
  \bibinfo {author} {\bibfnamefont {X.-L.}\ \bibnamefont {Qi}},\ }\href
  {http://arxiv.org/abs/1212.6951} {\bibfield  {journal} {\bibinfo  {journal}
  {\prb}\ }\textbf {\bibinfo {volume} {88}},\ \bibinfo {pages} {195412}
  (\bibinfo {year} {2013})},\ \Eprint {http://arxiv.org/abs/arXiv:1212.6951}
  {arXiv:1212.6951} \BibitemShut {NoStop}%
\bibitem [{\citenamefont {{Pollmann}}\ and\ \citenamefont
  {{Turner}}(2012)}]{PT1241}%
  \BibitemOpen
  \bibfield  {author} {\bibinfo {author} {\bibfnamefont {F.}~\bibnamefont
  {{Pollmann}}}\ and\ \bibinfo {author} {\bibfnamefont {A.~M.}\ \bibnamefont
  {{Turner}}},\ }\href {\doibase 10.1103/PhysRevB.86.125441} {\bibfield
  {journal} {\bibinfo  {journal} {\prb}\ }\textbf {\bibinfo {volume} {86}},\
  \bibinfo {pages} {125441} (\bibinfo {year} {2012})},\ \Eprint
  {http://arxiv.org/abs/1204.0704} {arXiv:1204.0704} \BibitemShut {NoStop}%
\bibitem [{\citenamefont {Hung}\ and\ \citenamefont {Wen}(2013)}]{HW1339}%
  \BibitemOpen
  \bibfield  {author} {\bibinfo {author} {\bibfnamefont {L.-Y.}\ \bibnamefont
  {Hung}}\ and\ \bibinfo {author} {\bibfnamefont {X.-G.}\ \bibnamefont {Wen}},\
  }\href@noop {} {\  (\bibinfo {year} {2013})},\ \Eprint
  {http://arxiv.org/abs/arXiv:1311.5539} {arXiv:1311.5539} \BibitemShut
  {NoStop}%
\bibitem [{\citenamefont {Wen}(2014{\natexlab{b}})}]{W1447}%
  \BibitemOpen
  \bibfield  {author} {\bibinfo {author} {\bibfnamefont {X.-G.}\ \bibnamefont
  {Wen}},\ }\href {\doibase 10.1103/PhysRevB.89.035147} {\bibfield  {journal}
  {\bibinfo  {journal} {Phys. Rev. B}\ }\textbf {\bibinfo {volume} {89}},\
  \bibinfo {pages} {035147} (\bibinfo {year} {2014}{\natexlab{b}})},\ \Eprint
  {http://arxiv.org/abs/arXiv:1301.7675} {arXiv:1301.7675} \BibitemShut
  {NoStop}%
\bibitem [{\citenamefont {Moradi}\ and\ \citenamefont {Wen}(2014)}]{MW1418}%
  \BibitemOpen
  \bibfield  {author} {\bibinfo {author} {\bibfnamefont {H.}~\bibnamefont
  {Moradi}}\ and\ \bibinfo {author} {\bibfnamefont {X.-G.}\ \bibnamefont
  {Wen}},\ }\href@noop {} {\  (\bibinfo {year} {2014})},\ \Eprint
  {http://arxiv.org/abs/arXiv:1401.0518} {arXiv:1401.0518} \BibitemShut
  {NoStop}%
\bibitem [{\citenamefont {He}\ \emph {et~al.}(2014)\citenamefont {He},
  \citenamefont {Moradi},\ and\ \citenamefont {Wen}}]{HMW1457}%
  \BibitemOpen
  \bibfield  {author} {\bibinfo {author} {\bibfnamefont {H.}~\bibnamefont
  {He}}, \bibinfo {author} {\bibfnamefont {H.}~\bibnamefont {Moradi}}, \ and\
  \bibinfo {author} {\bibfnamefont {X.-G.}\ \bibnamefont {Wen}},\ }\href@noop
  {} {\  (\bibinfo {year} {2014})},\ \Eprint
  {http://arxiv.org/abs/arXiv:1401.5557} {arXiv:1401.5557} \BibitemShut
  {NoStop}%
\bibitem [{\citenamefont {Gu}\ \emph {et~al.}(2008)\citenamefont {Gu},
  \citenamefont {Levin},\ and\ \citenamefont {Wen}}]{Gu_TnsrRG_08}%
  \BibitemOpen
  \bibfield  {author} {\bibinfo {author} {\bibfnamefont {Z.-C.}\ \bibnamefont
  {Gu}}, \bibinfo {author} {\bibfnamefont {M.}~\bibnamefont {Levin}}, \ and\
  \bibinfo {author} {\bibfnamefont {X.-G.}\ \bibnamefont {Wen}},\ }\href@noop
  {} {\bibfield  {journal} {\bibinfo  {journal} {Phys. Rev. B}\ }\textbf
  {\bibinfo {volume} {78}},\ \bibinfo {pages} {205116} (\bibinfo {year}
  {2008})}\BibitemShut {NoStop}%
\bibitem [{\citenamefont {Vidal}(2007{\natexlab{a}})}]{Vidal_1D_07}%
  \BibitemOpen
  \bibfield  {author} {\bibinfo {author} {\bibfnamefont {G.}~\bibnamefont
  {Vidal}},\ }\href@noop {} {\bibfield  {journal} {\bibinfo  {journal} {Phys.
  Rev. Lett.}\ }\textbf {\bibinfo {volume} {98}},\ \bibinfo {pages} {070201}
  (\bibinfo {year} {2007}{\natexlab{a}})}\BibitemShut {NoStop}%
\bibitem [{\citenamefont {Banuls}\ \emph {et~al.}(2009)\citenamefont {Banuls},
  \citenamefont {Hastings}, \citenamefont {Verstraete},\ and\ \citenamefont
  {Cirac}}]{Cirac_1D_09}%
  \BibitemOpen
  \bibfield  {author} {\bibinfo {author} {\bibfnamefont {M.~C.}\ \bibnamefont
  {Banuls}}, \bibinfo {author} {\bibfnamefont {M.~B.}\ \bibnamefont
  {Hastings}}, \bibinfo {author} {\bibfnamefont {F.}~\bibnamefont
  {Verstraete}}, \ and\ \bibinfo {author} {\bibfnamefont {J.~I.}\ \bibnamefont
  {Cirac}},\ }\href@noop {} {\bibfield  {journal} {\bibinfo  {journal} {Phys.
  Rev. Lett.}\ }\textbf {\bibinfo {volume} {102}},\ \bibinfo {pages} {240603}
  (\bibinfo {year} {2009})}\BibitemShut {NoStop}%
\bibitem [{\citenamefont {Perez-Garcia}\ \emph {et~al.}(2008)\citenamefont
  {Perez-Garcia}, \citenamefont {Wolf}, \citenamefont {Sanz}, \citenamefont
  {Verstraete},\ and\ \citenamefont {Cirac}}]{Cirac08}%
  \BibitemOpen
  \bibfield  {author} {\bibinfo {author} {\bibfnamefont {D.}~\bibnamefont
  {Perez-Garcia}}, \bibinfo {author} {\bibfnamefont {M.~M.}\ \bibnamefont
  {Wolf}}, \bibinfo {author} {\bibfnamefont {M.}~\bibnamefont {Sanz}}, \bibinfo
  {author} {\bibfnamefont {F.}~\bibnamefont {Verstraete}}, \ and\ \bibinfo
  {author} {\bibfnamefont {J.~I.}\ \bibnamefont {Cirac}},\ }\href@noop {}
  {\bibfield  {journal} {\bibinfo  {journal} {Phys. Rev. Lett.}\ }\textbf
  {\bibinfo {volume} {100}},\ \bibinfo {pages} {167202} (\bibinfo {year}
  {2008})}\BibitemShut {NoStop}%
\bibitem [{\citenamefont {Pollmann}\ \emph {et~al.}(2010)\citenamefont
  {Pollmann}, \citenamefont {Berg}, \citenamefont {Turner},\ and\ \citenamefont
  {Oshikawa}}]{PBT1039}%
  \BibitemOpen
  \bibfield  {author} {\bibinfo {author} {\bibfnamefont {F.}~\bibnamefont
  {Pollmann}}, \bibinfo {author} {\bibfnamefont {E.}~\bibnamefont {Berg}},
  \bibinfo {author} {\bibfnamefont {A.~M.}\ \bibnamefont {Turner}}, \ and\
  \bibinfo {author} {\bibfnamefont {M.}~\bibnamefont {Oshikawa}},\ }\href
  {\doibase 10.1103/PhysRevB.81.064439} {\bibfield  {journal} {\bibinfo
  {journal} {Phys. Rev. B}\ }\textbf {\bibinfo {volume} {81}},\ \bibinfo
  {pages} {064439} (\bibinfo {year} {2010})},\ \Eprint
  {http://arxiv.org/abs/arXiv:0910.1811} {arXiv:0910.1811} \BibitemShut
  {NoStop}%
\bibitem [{\citenamefont {Chen}\ \emph {et~al.}(2011)\citenamefont {Chen},
  \citenamefont {Gu},\ and\ \citenamefont {Wen}}]{CGW1107}%
  \BibitemOpen
  \bibfield  {author} {\bibinfo {author} {\bibfnamefont {X.}~\bibnamefont
  {Chen}}, \bibinfo {author} {\bibfnamefont {Z.-C.}\ \bibnamefont {Gu}}, \ and\
  \bibinfo {author} {\bibfnamefont {X.-G.}\ \bibnamefont {Wen}},\ }\href@noop
  {} {\bibfield  {journal} {\bibinfo  {journal} {Phys. Rev. B}\ }\textbf
  {\bibinfo {volume} {83}},\ \bibinfo {pages} {035107} (\bibinfo {year}
  {2011})},\ \Eprint {http://arxiv.org/abs/arXiv:1008.3745} {arXiv:1008.3745}
  \BibitemShut {NoStop}%
\bibitem [{\citenamefont {Schuch}\ \emph {et~al.}(2011)\citenamefont {Schuch},
  \citenamefont {Perez-Garcia},\ and\ \citenamefont {Cirac}}]{SPC1139}%
  \BibitemOpen
  \bibfield  {author} {\bibinfo {author} {\bibfnamefont {N.}~\bibnamefont
  {Schuch}}, \bibinfo {author} {\bibfnamefont {D.}~\bibnamefont
  {Perez-Garcia}}, \ and\ \bibinfo {author} {\bibfnamefont {I.}~\bibnamefont
  {Cirac}},\ }\href@noop {} {\bibfield  {journal} {\bibinfo  {journal} {Phys.
  Rev. B}\ }\textbf {\bibinfo {volume} {84}},\ \bibinfo {pages} {165139}
  (\bibinfo {year} {2011})},\ \Eprint {http://arxiv.org/abs/arXiv:1010.3732}
  {arXiv:1010.3732} \BibitemShut {NoStop}%
\bibitem [{\citenamefont {Vidal}(2007{\natexlab{b}})}]{V0701}%
  \BibitemOpen
  \bibfield  {author} {\bibinfo {author} {\bibfnamefont {G.}~\bibnamefont
  {Vidal}},\ }\href {\doibase 10.1103/PhysRevLett.98.070201} {\bibfield
  {journal} {\bibinfo  {journal} {Phys. Rev. Lett.}\ }\textbf {\bibinfo
  {volume} {98}},\ \bibinfo {pages} {070201} (\bibinfo {year}
  {2007}{\natexlab{b}})},\ \Eprint {http://arxiv.org/abs/cond-mat/0605597}
  {cond-mat/0605597} \BibitemShut {NoStop}%
\bibitem [{\citenamefont {Xiang}\ \emph {et~al.}(2001)\citenamefont {Xiang},
  \citenamefont {Lou},\ and\ \citenamefont {Su}}]{Tao_DMRG_01}%
  \BibitemOpen
  \bibfield  {author} {\bibinfo {author} {\bibfnamefont {T.}~\bibnamefont
  {Xiang}}, \bibinfo {author} {\bibfnamefont {J.}~\bibnamefont {Lou}}, \ and\
  \bibinfo {author} {\bibfnamefont {Z.}~\bibnamefont {Su}},\ }\href@noop {}
  {\bibfield  {journal} {\bibinfo  {journal} {Phys. Rev. B}\ }\textbf {\bibinfo
  {volume} {64}},\ \bibinfo {pages} {104414} (\bibinfo {year}
  {2001})}\BibitemShut {NoStop}%
\bibitem [{\citenamefont {Legeza}\ and\ \citenamefont
  {Solyom}(2004)}]{Legeza_DMRG_04}%
  \BibitemOpen
  \bibfield  {author} {\bibinfo {author} {\bibfnamefont {O.}~\bibnamefont
  {Legeza}}\ and\ \bibinfo {author} {\bibfnamefont {J.}~\bibnamefont
  {Solyom}},\ }\href@noop {} {\bibfield  {journal} {\bibinfo  {journal} {Phys.
  Rev. B}\ }\textbf {\bibinfo {volume} {70}},\ \bibinfo {pages} {205118}
  (\bibinfo {year} {2004})}\BibitemShut {NoStop}%
\bibitem [{\citenamefont {Vidal}\ \emph {et~al.}(2003)\citenamefont {Vidal},
  \citenamefont {Latorre}, \citenamefont {Rico},\ and\ \citenamefont
  {Kitaev}}]{Vidal_AreaLaw_03}%
  \BibitemOpen
  \bibfield  {author} {\bibinfo {author} {\bibfnamefont {G.}~\bibnamefont
  {Vidal}}, \bibinfo {author} {\bibfnamefont {J.~I.}\ \bibnamefont {Latorre}},
  \bibinfo {author} {\bibfnamefont {E.}~\bibnamefont {Rico}}, \ and\ \bibinfo
  {author} {\bibfnamefont {A.}~\bibnamefont {Kitaev}},\ }\href@noop {}
  {\bibfield  {journal} {\bibinfo  {journal} {Phys. Rev. Lett.}\ }\textbf
  {\bibinfo {volume} {90}},\ \bibinfo {pages} {227902} (\bibinfo {year}
  {2003})}\BibitemShut {NoStop}%
\bibitem [{\citenamefont {Srednicki}(1993)}]{Srednicki_93}%
  \BibitemOpen
  \bibfield  {author} {\bibinfo {author} {\bibfnamefont {M.}~\bibnamefont
  {Srednicki}},\ }\href@noop {} {\bibfield  {journal} {\bibinfo  {journal}
  {Phys. Rev. Lett.}\ }\textbf {\bibinfo {volume} {71}},\ \bibinfo {pages}
  {666} (\bibinfo {year} {1993})}\BibitemShut {NoStop}%
\bibitem [{\citenamefont {Plenio}\ \emph {et~al.}(2005)\citenamefont {Plenio},
  \citenamefont {Eisert}, \citenamefont {DreiBig},\ and\ \citenamefont
  {Cramer}}]{Plenio_AreaLaw_05}%
  \BibitemOpen
  \bibfield  {author} {\bibinfo {author} {\bibfnamefont {M.~B.}\ \bibnamefont
  {Plenio}}, \bibinfo {author} {\bibfnamefont {J.}~\bibnamefont {Eisert}},
  \bibinfo {author} {\bibfnamefont {J.}~\bibnamefont {DreiBig}}, \ and\
  \bibinfo {author} {\bibfnamefont {M.}~\bibnamefont {Cramer}},\ }\href@noop {}
  {\bibfield  {journal} {\bibinfo  {journal} {Phys. Rev. Lett.}\ }\textbf
  {\bibinfo {volume} {94}},\ \bibinfo {pages} {060503} (\bibinfo {year}
  {2005})}\BibitemShut {NoStop}%
\bibitem [{\citenamefont {Blote}\ and\ \citenamefont
  {Deng}(2002)}]{Deng_QMC_02}%
  \BibitemOpen
  \bibfield  {author} {\bibinfo {author} {\bibfnamefont {H.~W.~J.}\
  \bibnamefont {Blote}}\ and\ \bibinfo {author} {\bibfnamefont
  {Y.}~\bibnamefont {Deng}},\ }\href@noop {} {\bibfield  {journal} {\bibinfo
  {journal} {Phys. Rev. E}\ }\textbf {\bibinfo {volume} {66}},\ \bibinfo
  {pages} {066110} (\bibinfo {year} {2002})}\BibitemShut {NoStop}%
\bibitem [{\citenamefont {Gu}\ \emph {et~al.}(2009)\citenamefont {Gu},
  \citenamefont {Levin}, \citenamefont {Swingle},\ and\ \citenamefont
  {Wen}}]{Gu_TPS_09}%
  \BibitemOpen
  \bibfield  {author} {\bibinfo {author} {\bibfnamefont {Z.-C.}\ \bibnamefont
  {Gu}}, \bibinfo {author} {\bibfnamefont {M.}~\bibnamefont {Levin}}, \bibinfo
  {author} {\bibfnamefont {B.}~\bibnamefont {Swingle}}, \ and\ \bibinfo
  {author} {\bibfnamefont {X.-G.}\ \bibnamefont {Wen}},\ }\href@noop {}
  {\bibfield  {journal} {\bibinfo  {journal} {Phys. Rev. B}\ }\textbf {\bibinfo
  {volume} {79}},\ \bibinfo {pages} {085118} (\bibinfo {year}
  {2009})}\BibitemShut {NoStop}%
\bibitem [{\citenamefont {Trebst}\ \emph {et~al.}(2007)\citenamefont {Trebst},
  \citenamefont {Werner}, \citenamefont {Troyer}, \citenamefont {Shtengel},\
  and\ \citenamefont {Nayak}}]{Nayak_TC_07}%
  \BibitemOpen
  \bibfield  {author} {\bibinfo {author} {\bibfnamefont {S.}~\bibnamefont
  {Trebst}}, \bibinfo {author} {\bibfnamefont {P.}~\bibnamefont {Werner}},
  \bibinfo {author} {\bibfnamefont {M.}~\bibnamefont {Troyer}}, \bibinfo
  {author} {\bibfnamefont {K.}~\bibnamefont {Shtengel}}, \ and\ \bibinfo
  {author} {\bibfnamefont {C.}~\bibnamefont {Nayak}},\ }\href@noop {}
  {\bibfield  {journal} {\bibinfo  {journal} {Phys. Rev. Lett.}\ }\textbf
  {\bibinfo {volume} {98}},\ \bibinfo {pages} {070602} (\bibinfo {year}
  {2007})}\BibitemShut {NoStop}%
\bibitem [{\citenamefont {Chen}\ \emph
  {et~al.}(2010{\natexlab{b}})\citenamefont {Chen}, \citenamefont {Zeng},
  \citenamefont {Gu}, \citenamefont {Chuang},\ and\ \citenamefont
  {Wen}}]{Xie_NSymmCond_10}%
  \BibitemOpen
  \bibfield  {author} {\bibinfo {author} {\bibfnamefont {X.}~\bibnamefont
  {Chen}}, \bibinfo {author} {\bibfnamefont {B.}~\bibnamefont {Zeng}}, \bibinfo
  {author} {\bibfnamefont {Z.-C.}\ \bibnamefont {Gu}}, \bibinfo {author}
  {\bibfnamefont {I.~L.}\ \bibnamefont {Chuang}}, \ and\ \bibinfo {author}
  {\bibfnamefont {X.-G.}\ \bibnamefont {Wen}},\ }\href@noop {} {\bibfield
  {journal} {\bibinfo  {journal} {Phys. Rev. B}\ }\textbf {\bibinfo {volume}
  {82}},\ \bibinfo {pages} {165119} (\bibinfo {year}
  {2010}{\natexlab{b}})}\BibitemShut {NoStop}%
\end{thebibliography}%

\end{document}